\pdfoutput=1
\newcommand*{\ATLASLATEXPATH}{latex/}
\documentclass[UKenglish, texlive=2016, cernpreprint]{\ATLASLATEXPATH atlasdoc}
\LEcontact{Chip Brock brock@pa.msu.edu}

\usepackage{\ATLASLATEXPATH atlaspackage}
\usepackage{mathtools}
\usepackage{\ATLASLATEXPATH atlasbiblatex}

\usepackage{\ATLASLATEXPATH atlasphysics}
\graphicspath{{logos/}{figures/}}

\usepackage{hh4b-2017-paper-defs}

\AtlasTitle{
Search for pair production of Higgs bosons in the \bbbb final state using proton-proton collisions at $\sqrt{s} = 13$ TeV with the ATLAS detector
}

\AtlasAbstract{A search for Higgs boson pair production in the $b\bar{b}b\bar{b}$ final state is carried out with up to 36.1 $\mathrm{fb}^{-1}$ of LHC proton--proton collision data collected at $\sqrt{s}$ = 13 TeV with the ATLAS detector in 2015 and 2016. Three benchmark signals are studied: a spin-2 graviton decaying into a Higgs boson pair, a scalar resonance decaying into a Higgs boson pair, and Standard Model non-resonant Higgs boson pair production. Two analyses are carried out, each implementing a particular technique for the event reconstruction that targets Higgs bosons reconstructed as pairs of jets or single boosted jets. The resonance mass range covered is 260--3000 GeV. The analyses are statistically combined and upper limits on the production cross section of Higgs boson pairs times branching ratio to $b\bar{b}b\bar{b}$ are set in each model. No significant excess is observed; the largest deviation of data over prediction is found at a mass of 280 GeV, corresponding to 2.3 standard deviations globally. The observed 95\% confidence level upper limit on the non-resonant production is 13 times the Standard Model prediction.}

\author{The ATLAS Collaboration}

\AtlasRefCode{EXOT-2016-31}

\PreprintIdNumber{CERN-EP-2018-029}

\HepDataRecord{https://www.hepdata.net/record/82599}

\AtlasJournal{JHEP}
\AtlasJournalRef{JHEP 01 (2019) 030}
\AtlasDOI{10.1007/JHEP01(2019)030}

\AtlasCoverSupportingNote{Resolved support}{https://cds.cern.ch/record/2270210}
\AtlasCoverSupportingNote{Boosted support}{https://cds.cern.ch/record/2270230}

\AtlasCoverAnalysisTeam{D.~Abbott, J.~Alison, O.~Brandt, P.~Bryant, S.~Dong, M.~Franklin, J.~Frost, A.~Goussiou, K.~Gregersen, C.~Issever, M.~Kagan, L.~Kaplan, N.~Konstantinidis, T.~Lazovich, T.~Lenz, A.~Melzer, M.~Montella, N.~Norjoharuddeen, M.~Proffitt, G.~Putnam, M.~Sahinsoy, J.~Schaarschmidt, M.~Shochet, G.~Rosin, B.~Tong, A.~Tuna, D.~Wardrope, S.~Willocq, S.~Wu, Q.~Zeng}

\AtlasCoverEdBoardMember{Andrew Mehta~(chair)}
\AtlasCoverEdBoardMember{Georges Aad}
\AtlasCoverEdBoardMember{Chris Pollard}

\AtlasCoverEgroupEditors{atlas-exot-2016-31-editors@cern.ch}

\AtlasCoverEgroupEdBoard{atlas-exot-2016-31-editorial-board@cern.ch}

\hypersetup{pdftitle={ATLAS document},pdfauthor={The ATLAS Collaboration}}

\begin{document}
\maketitle
\section{Introduction}
\label{sec:intro}

The discovery of the Standard Model (SM) Higgs boson (\h) \cite{Aad:2012tfa,Chatrchyan201230} at the Large Hadron Collider (LHC) motivates searches for new physics using the Higgs boson as a probe. In particular, many models predict cross sections for Higgs boson pair production that are significantly greater than the SM prediction. Resonant Higgs boson pair production is predicted by models such as the bulk Randall--Sundrum model~\cite{Agashe:2007zd,Fitzpatrick}, which features spin-2 Kaluza--Klein gravitons, \Grav, that subsequently decay to pairs of Higgs bosons. Extensions of the Higgs sector, such as two-Higgs-doublet models~\cite{PhysRevD.8.1226, Branco:2011iw}, propose the existence of a heavy spin-0 scalar that can decay into \h pairs. Enhanced non-resonant Higgs boson pair production is predicted by other models, for example those featuring light coloured scalars \cite{PhysRevD.86.095023} or direct $t\bar{t}\hh$ vertices \cite{Grober:2010yv,Contino:2012xk}.

Previous searches for Higgs boson pair production have all yielded null results. In the \bbbb channel, ATLAS searched for both non-resonant and resonant production in the mass range 400--3000~\GeV\ using 3.2 $\mathrm{fb}^{-1}$ of 13~\TeV\ data~\cite{EXOT-2015-11} collected during 2015. CMS searched for the production of resonances with masses 750--3000~\GeV\ using 13~\TeV\ data~\cite{CMS-B2G-16-026} and with masses 270--1100~\GeV\ with 8~\TeV\ data~\cite{CMS-HIG-14-013}. Using 8~\TeV\ data, ATLAS has examined the \bbbb~\cite{Aad:2015uka}, \bbgg~\cite{HIGG-2013-29}, \bbtautau and \WWgg channels, all of which were combined in Ref.~\cite{HIGG-2013-33}. CMS has performed searches using 13~\TeV\ data for the \bbtautau~\cite{CMSHIG17002} and $bb\ell\nu\ell\nu$~\cite{CMSHIG17006} final states, and used 8~\TeV\ data to search for \bbgg~\cite{Khachatryan:2016sey} in addition to a search in multilepton and multilepton+photons final states~\cite{Khachatryan:2014jya}.

The analyses presented in this paper exploit the decay mode with the largest branching ratio, \hbb, to search for Higgs boson pair production in both resonant and non-resonant production. Two analyses, which are complementary in their acceptance, are presented, each employing a unique technique to reconstruct the Higgs bosons. The resolved analysis is used for \hh systems in which the Higgs bosons have Lorentz boosts low enough that four $b$-jets can be reconstructed. The boosted analysis is used for those \hh systems in which the Higgs bosons have higher Lorentz boosts, which prevents the Higgs boson decay products from being resolved in the detector as separate \bjets. Instead, each Higgs boson candidate consists of a single large-radius jet, and the presence of $b$-quarks is inferred using smaller-radius jets built from charged-particle tracks.

Both analyses were re-optimized with respect to the previous ATLAS publication~\cite{EXOT-2015-11}; an improved algorithm to pair $b$-jets to Higgs boson candidates is used in the resolved analysis, and in the boosted analysis an additional signal-enriched sample is utilized. The dataset comprises 2015 and 2016 data, corresponding to 27.5 $\mathrm{fb}^{-1}$ for the resolved analysis and 36.1 $\mathrm{fb}^{-1}$ for the boosted analysis, with the difference due to the trigger selections used. The results are obtained using the resolved analysis for a resonance mass between 260 and 1400~\GeV, and the boosted analysis for masses between 800~\GeV\ and 3000~\GeV. The main background is multijet production, which is estimated from data; the sub-leading background is \ttbar, which is estimated using both data and simulations. The two analyses employ orthogonal selections, and a statistical combination is performed in the mass range where they overlap. The final discriminants are the four-jet and dijet mass distributions in the resolved and boosted analyses, respectively. Searches are performed for the following benchmark signals: a spin-2 graviton decaying into Higgs bosons, a scalar resonance decaying into a Higgs boson pair, and SM non-resonant Higgs boson pair production.

\clearpage

\section{ATLAS detector}
\label{sec:detector}
\newcommand{\AtlasCoordFootnote}{
ATLAS uses a right-handed coordinate system with its origin at the nominal interaction point (IP) in the centre of the detector and the $z$-axis along the beam pipe. The $x$-axis points from the IP to the centre of the LHC ring, and the $y$-axis points upwards. Cylindrical coordinates $(r,\phi)$ are used in the transverse plane, $\phi$ being the azimuthal angle around the $z$-axis. The pseudorapidity is defined in terms of the polar angle $\theta$ as $\eta = -\ln \tan(\theta/2)$. Angular distance is measured in units of $\Delta R \equiv \sqrt{(\Delta\eta)^{2} + (\Delta\phi)^{2}}$.}

The ATLAS experiment~\cite{PERF-2007-01} at the LHC is a multipurpose particle detector with a forward-backward symmetric cylindrical geometry and a near $4\pi$ coverage in solid angle.\footnote{\AtlasCoordFootnote} It consists of an inner tracking detector (ID) surrounded by a thin superconducting solenoid providing a \SI{2}{\tesla} axial magnetic field, electromagnetic (EM) and hadronic calorimeters, and a muon spectrometer (MS). The ID covers the pseudorapidity range $|\eta| < 2.5$. It consists of silicon pixel, silicon microstrip, and straw-tube transition-radiation tracking detectors. An additional pixel detector layer~\cite{Capeans:1291633}, inserted at a mean radius of 3.3 cm, improves the identification of $b$-jets~\cite{ATL-PHYS-PUB-2015-022}. Lead/liquid-argon (LAr) sampling calorimeters provide EM energy measurements. A steel/scintillator-tile hadronic calorimeter covers the central pseudorapidity range ($|\eta| < 1.7$). The endcap and forward regions are instrumented with LAr calorimeters for both the EM and hadronic energy measurements up to $|\eta| = 4.9$. The muon spectrometer surrounds the calorimeters and includes three large superconducting air-core toroids. The field integral of the toroids ranges between \num{2.0} and \SI{6.0}{\tesla\metre} for most of the detector. The MS includes a system of precision tracking chambers and triggering chambers. A dedicated trigger system is used to select events~\cite{atlas-trigger}. The first-level trigger is implemented in hardware and uses the calorimeter and muon detectors to reduce the accepted event rate to \SI{100}{\kilo\hertz}. This is followed by a software-based high-level trigger that reduces the accepted event rate to \SI{1}{\kilo\hertz} on average.

\section{Data and simulation}
\label{sec:dataMC}

\subsection{Data}

This analysis is performed on two LHC $pp$ collision datasets at $\sqrt{s} = 13$~\TeV. Data were collected during stable beam conditions and when all relevant detector systems were functional. The integrated luminosity of the dataset collected during 2015 was measured to be 3.2~\ifb.  The second dataset was collected during 2016 and corresponds to an integrated luminosity of 24.3~\ifb\ for the resolved analysis and 32.9~\ifb\ for the boosted analysis.

The difference in integrated luminosity between the two analyses results from the choices of triggers. In the resolved analysis, a combination of $b$-jet triggers is used. Events were required to feature either one $b$-tagged jet~\cite{btaggingRun2, Aad:2015ydr} with transverse momentum $\pt>225$~\GeV, or two $b$-tagged jets, either both satisfying $\pt>35$~\GeV\ or both satisfying $\pt>55$~\GeV, with different requirements on the $b$-tagging. Some triggers required additional non-$b$-tagged jets. Due to a change in the online $b$-tagging algorithm between 2015 and 2016, the two datasets are treated independently until they are combined in the final statistical analysis. After the selection described in Section~\ref{sec:resolved}, this combination of triggers is estimated to be 65\% efficient for simulated signals with a Higgs boson pair invariant mass, $m_{HH}$, of 280~\GeV, rising to 100\% efficiency for resonance masses greater than 600~\GeV. During 2016 data-taking, a fraction of the data was affected by an inefficiency in the online vertex reconstruction, which reduced the efficiency of the algorithms used to identify $b$-jets; those events were not retained for further analysis. This reduces the integrated luminosity of the 2016 dataset for the resolved analysis to 24.3 \ifb. In the boosted analysis, events were selected from the 2015 dataset using a trigger that required a single \akt jet~\cite{Cacciari:2008gp} with radius parameter $R=1.0$ and with $\pt > 360$~\GeV. In 2016, a similar trigger was used but with a higher threshold of $\pt > 420$~\GeV. The efficiency of these triggers is 100\% for simulated signals passing the jet requirements described in Section~\ref{sec:boosted}, so the 2015 and 2016 datasets were combined into one dataset. 

\subsection{Signal models and simulation}
Simulated Monte Carlo (MC) event samples are used in this analysis to model signal production and
the background from \ttbar. The dominant multi-jet background is modelled using data-driven techniques, as described in Sections~\ref{sec:resolvedBKG} and~\ref{sec:boostedBackground}.

$gg\to$ Scalar \tohhb\ events were generated at LO in QCD with \MGamcatnlo~2.2.3 interfaced with \textsc{Herwig++}~\cite{Bahr:2008pv} for parton-showering, hadronization and simulation of the underlying event. CT10~\cite{Lai:2010vv} PDF sets were used for \MGamcatnlo and CTEQ6L1~\cite{cteq6l1} for \textsc{Herwig++}. The UE-EE-5-CTEQ6L1 set of tuned underlying-event parameters \cite{Seymour:2013qka} was used. No specific model was considered for computing the scalar signal cross sections.

Signal \Gtohhb\ events were generated at leading order (LO) with \MGamcatnlo~2.2.2~\cite{MGaMCatNLO} interfaced with \PYTHIA~8.186~\cite{Sjostrand:2007gs} for parton-showering, hadronization and underlying-event simulation. The NNPDF2.3 LO
parton distribution function (PDF) set~\cite{Ball:2012cx} was used for both \MGamcatnlo and \PYTHIA. The A14 set of tuned underlying-event parameters~\cite{ATL-PHYS-PUB-2014-021} was used. These signal samples were generated with $\kMPl = 1$ or 2, where $k$ is the curvature of the warped extra dimension and $\overline{M}_{\mathrm{Pl}} = 2.4 \times 10^{18}$~\GeV\ is the effective four-dimensional Planck scale.

For the evaluation of theoretical uncertainties in the signal modeling, samples were produced with variations of the factorization and renormalization scales, PDF sets (following the prescription from Ref.~\cite{pdfs}) and shower generator. For the latter, scalar (spin-2) samples were produced that are interfaced to \PYTHIA~8.186 rather than \textsc{Herwig++} (and vice versa).

The decay widths of these three resonance models differ. The scalar signals were generated with a width of 1~\GeV, allowing a study of generic narrow-width scalar signals. The widths of the graviton signals depend on the resonance mass and the value of $\kMPl$.
Relative to the resonance mass, they range from 3\% (6\%) at low mass to 13\% (25\%) at the highest mass for $\kMPl=1$ ($\kMPl=2$).
The graviton samples were normalized using cross sections from Ref.~\cite{carvalho}.

Resonant signal samples for the scalar and $\kMPl=1$ graviton models were produced with masses in 10~\GeV\ steps between 260 and 300~\GeV, in 100~\GeV\ steps up to 1600~\GeV, in 200~\GeV\ steps up to 2000~\GeV, and in 250~\GeV\ steps up to 3000~\GeV. Signal samples for the $\kMPl=2$ graviton model were produced with the same spacings but omitting the masses of 270~\GeV, 290~\GeV\ and 2750~\GeV\ due to the larger generated width.

SM non-resonant production of Higgs boson pairs via the gluon--gluon fusion process was simulated at NLO with \MGamcatnlo~\cite{Frederix:2014hta}, using form factors for the top-quark loop from HPAIR~\cite{PhysRevD.58.115012, Plehn199646} to approximate finite top-quark mass effects. The simulated events were reweighted to reproduce the \mhh spectrum obtained in Refs.~\cite{Borowka:2016ehy, Borowka:2016ypz}, which calculated the process at NLO in QCD while fully accounting for the top-quark mass. The cross section times branching ratio to the \fourb\ final state, evaluated at next-to-next-to-leading order (NNLO) with the summation of logarithms at next-to-next-leading-logarithm (NNLL) accuracy and including top-quark mass effects at NLO is $11.3^{+0.9}_{-1.0}$~fb~{\cite{Borowka:2016ehy}. The uncertainty includes the effects due to renormalization and factorization scales, PDF set, $\alpha_{\mathrm{S}}$, and the \hbb\ branching ratio. In all signal samples, the mass of the Higgs boson ($m_H$) was set to 125~\GeV.

Interference effects between \HH resonant production and SM non-resonant \HH production are not included in the simulated samples.

The generation of \ttbar\ events was performed with \powhegbox\ v1~\cite{Alioli:2010xd} using the CT10 PDF set. The parton shower, hadronization, and the underlying event were simulated using \PYTHIA~6.428~\cite{Sjostrand:2006za} with the CTEQ6L1 PDF set and the corresponding Perugia 2012 set of tuned underlying-event parameters~\cite{Skands:2010ak}. The top-quark mass was set to 172.5~\GeV{}. Higher-order corrections to the \ttbar\ cross section were computed with Top++~2.0~\cite{Czakon:20142930}. These incorporate NNLO corrections in QCD, including resummation of NNLL soft gluon terms.

The $Z$+jets sample was generated using \PYTHIA~8.186 with the NNPDF2.3 LO PDF set.

For all simulated samples, charm-hadron and bottom-hadron decays were handled by {\textsc{EvtGen}} 1.2.0~\cite{EvtGen}. To simulate the impact of multiple $pp$ interactions that occur within the same or nearby bunch crossings (pile-up), minimum-bias events generated with \PYTHIA~8 using the A2 set of tuned parameters~\cite{MC12AU2} were overlaid on the hard-scatter event. The detector response was simulated with \GEANT~4~\cite{Agostinelli:2002hh, Aad:2010ah} and the events were processed with the same reconstruction software as that used for the data. 

\section{Object reconstruction}
\label{sec:reco}
Jets are built from topological clusters of energy deposits in calorimeter cells \cite{PERF-2014-07}, using a four-momentum reconstruction scheme with massless clusters as input. The directions of jets are corrected to point back to the identified hard-scatter, proton--proton collision vertex, which is the vertex with the highest $\Sigma \pt^2$ of constituent tracks.

Jets are reconstructed using the \akt algorithm with different values of the radius parameter \R. The jets with $R=0.4$ (``small-\R jets''), used in the resolved analysis, are reconstructed from clusters calibrated at the electromagnetic (EM) scale~\cite{Aad:2011he}. The jets are corrected for additional energy deposited from pile-up interactions using an area-based correction \cite{Cacciari:2008gn}. They are then calibrated using \pt- and $\eta$-dependent calibration factors derived from simulation, before global sequential calibration~\cite{Aad:2011he} is applied, which reduces differences in calorimeter responses to gluon- or quark-initiated jets. The final calibration is based on in situ measurements in collision data~\cite{ATL-PHYS-PUB-2015-015}. Jets with $\pT<60$~\GeV, $|\eta|<2.4$, and with a large fraction of their energy arising from pile-up interactions are suppressed using tracking information, which was combined in a multivariate classification algorithm (\emph{jet vertex tagger})~\cite{ATL-PHYS-PUB-2014-001}. Events that pass a ``medium'' jet vertex tagger working point, corresponding to a 92\% efficiency for jets at the EM scale with $20<\pt<60$~\GeV, are retained in the analysis. Quality criteria are applied to the jets, and events with jets consistent with noise in the calorimeter or non-collision backgrounds are vetoed~\cite{jetcleanATLAS}. 

The jets with $R=1.0$ (``large-\R jets'') used in the boosted analysis are built from locally calibrated~\cite{Aad:2011he} topological clusters. They are trimmed~\cite{Krohn2010} to minimize the impact of energy deposits from pile-up interactions. Trimming proceeds by reclustering the jet with the \kt algorithm~\cite{Ellis:1993tq} into $R = 0.2$ subjets and then removing those subjets with $\pt^{\mathrm{subjet}}/\pt^{\mathrm{jet}} < 0.05$, where $\pt^{\mathrm{subjet}}$ is the transverse momentum of the subjet and $\pt^{\mathrm{jet}}$ that of the original jet. The energy and mass scales of the trimmed jets are then calibrated using \pt- and $\eta$-dependent calibration factors derived from simulation~\cite{PERF-2012-02}. The mass of the large-\R jets is computed using tracking and calorimeter information, also called the \emph{combined mass} technique~\cite{ATLAS-CONF-2016-035}, which leads to a smaller mass resolution and better estimate of the median mass value than obtained using only calorimeter energy clusters.

Jets containing $b$-hadrons are identified using a score value computed from a multivariate $b$-tagging algorithm \emph{MV2c10}~\cite{btaggingRun2, Aad:2015ydr}, which makes use of observables provided by an impact parameter algorithm, an inclusive secondary vertex finding algorithm and a multi-vertex finding algorithm. The \emph{MV2c10} algorithm is applied to a set of charged-particle tracks that satisfy quality and impact parameter criteria and are matched to each jet. For large-\R jets, $b$-tagging is performed on \akt $R=0.2$ track-jets~\cite{ATL-PHYS-PUB-2014-013} matched to the large-\R jets using ghost association~\cite{Cacciari:2007fd}. These track-jets are required to be consistent with the primary vertex of the event as well as to satisfy \pt\ $>$ 10 \GeV{} and $|\eta| <$ 2.5. The small radius parameter of the track-jets enables two nearby $b$-hadrons to be identified when their \DR separation is less than 0.4, which is beneficial when reconstructing high-\pt Higgs boson candidates. The $b$-tagging requirements of both the resolved and the boosted analyses use working points that lead to an efficiency of 70\% for $b$-jets with \pt > 20~\GeV\ when evaluated in a sample of simulated \ttbar events. This working point corresponds to a rejection rate of jets originating from $u$-, $d$- or $s$-quarks or gluons of 380 for the jets with $R=0.4$ and 120 for the track-jets.  The rejection of jets from $c$-quarks is 12 (7.1) for the $R=0.4$ jets (track-jets).

Muons are reconstructed by combining tracks in the ID with tracks in the MS, and are required to have $\pt > 4~\GeV$, $|\eta| < 2.5$ and to satisfy ``medium'' muon identification criteria~\cite{PERF-2015-10}. If a muon is within $\DR = 0.4$ (0.2) of a jet used for $b$-tagging in the resolved (boosted) analysis, their four-momentum is added to the calorimeter-based jet's four-momentum to partially account for the energy lost in semileptonic $b$-hadron decays. 

\section{Resolved analysis}
\label{sec:resolved}
The resolved analysis is optimized to discover signals that result in either non-resonant or low-mass resonant Higgs boson pair production. The strategy is to select two Higgs boson candidates, each composed of two $b$-tagged \akt small-\R jets, with invariant masses near $m_H$. 

The invariant mass of the two-Higgs-boson-candidate system (\mfourj) is used as the final discriminant between Higgs boson pair production and the backgrounds (which are principally multijet, with some \ttbar). Resonant signals would lead to a localized excess, while non-resonant production would result in an excess in the tail of the \mfourj spectrum.

\subsection{Selection}
\label{sec:resolvedSelection}

Events are required to have a primary vertex with at least two tracks matched to it. The selection proceeds with the requirement that the event contains at least four $b$-tagged \akt small-\R jets with $\pt>40$~\GeV\ and $|\eta|<2.5$ (``four-tag'' sample).
The four jets with the highest $b$-tagging score are paired to construct two Higgs boson candidates. Initially, all possible pairings are considered. The angle between the decay products of the Higgs boson in the laboratory frame depends on the value of \mfourj and thus on the Lorentz boost of the Higgs boson. Accordingly, pairings of jets into Higgs boson candidates are only accepted if they satisfy the following requirements: 
\[
\begin{drcases}
\frac{360~\GeV}{\mfourj} - 0.5 < \DR_{jj, \mathrm{lead}} < \frac{653~\GeV}{\mfourj} + 0.475\\
\frac{235~\GeV}{\mfourj}\qquad\ < \DR_{jj, \mathrm{subl}} < \frac{875~\GeV}{\mfourj} + 0.35
\end{drcases}
\mathrm{if}\ \mfourj < 1250~\GeV,
\]
\[
\begin{drcases}
0< \DR_{jj, \mathrm{lead}} < 1\\
0 < \DR_{jj, \mathrm{subl}} < 1
\end{drcases}
\mathrm{if}\ \mfourj > 1250~\GeV.
\]
In these expressions, $\DR_{jj, \mathrm{lead}}$ is the angular distance between the jets in the leading Higgs boson candidate and $\DR_{jj, \mathrm{subl}}$ for the sub-leading candidate. The leading Higgs boson candidate is defined as the candidate with the highest scalar sum of jet \pt. These requirements on $\DR_{jj}$ efficiently reject jet-pairings in which one of the $b$-tagged jets is not consistent with one originating from a Higgs boson decay. The specific numbers in this and the following selection requirements were chosen to maximize the sensitivity to the signal.

Events that have more than two Higgs boson candidates satisfying these requirements necessitate an algorithm in order to choose the correct pairs. In the absence of energy loss through semileptonic $B$-decays, the optimal choice would be the combination most consistent with the decays of two particles of equal mass.\footnote{Explicitly requiring the masses to be equal to 125~\GeV\ does not significantly increase signal efficiency, while it sculpts the background Higgs boson candidates' mass distributions to look like those of the signal, reducing the signal vs background discrimination in these variables.}

To account for energy loss, the requirement of equal masses is modified. The distance, \Dhh, of the pairing's leading and sub-leading Higgs boson candidate's masses, $\left(\leadm, \sublm\right)$ from the line connecting $\left(0~\GeV, 0~\GeV\right)$ and $\left(120~\GeV, 110~\GeV\right)$ is computed, and the pairing with the smallest value of \Dhh is chosen. The values of 120~\GeV\ and 110~\GeV\ are chosen because they correspond to the median values of the narrowest intervals that contain 90\% of the signal in simulation. The quantity \Dhh can be expressed as follows:
\begin{equation*}
\Dhh = \frac{\left|\leadm - \frac{120}{110}\sublm\right|}{\sqrt{1+\left(\frac{120}{110}\right)^{2}}}.
\end{equation*}

In signal simulation the pairing of jets (when four $b$-jets have been identified) is correct in at least 90\% of the events, depending on $m_{4j}$.

The resolved analysis searches for resonances with a wide range of masses, 260~\GeV$<\mhh<1400$~\GeV, as well as non-resonant signals. Event selection criteria that vary as a function of \mfourj\ are used to reject background and hence enhance the analysis sensitivity across this range. Mass-dependent requirements are imposed on the \pt\ of the leading Higgs boson candidate, \ptlead, and the sub-leading Higgs boson candidate, \ptsubl:
\[
\ptlead > 0.5\mfourj - 103~\GeV,
\]
\[
\ptsubl > 0.33\mfourj - 73~\GeV.
\]

A further (\mfourj-independent) requirement is placed on the pseudorapidity difference between the two Higgs boson candidates, $\dEta < 1.5$, which rejects multijet events.

A requirement on the Higgs boson candidates' masses is used to define the signal region:
\begin{equation}
\Xhh = \sqrt{\left(\frac{\mlead - 120~\GeV}{0.1\mlead}\right)^2 + \left(\frac{\msubl - 110~\GeV}{0.1\msubl}\right)^2} < 1.6,
\label{eqn:resolvedXhh}
\end{equation}
where the 0.1\mtwoj terms represent the widths of the leading and sub-leading Higgs boson candidates' mass distributions, derived from simulation. The signal region is shown as the inner region of Figure~\ref{fig:resolvedRegions}.
 
To reduce the \ttbar background, hadronically decaying top-quark candidates are built from any three jets in the event, of which one must be a constituent of a Higgs boson candidate. These three jets are ordered by their $b$-tagging score. The highest one is considered as the $b$-jet originating from the top-quark candidate decay; the other two jets are considered as forming a hadronically decaying $W$ boson candidate. A measure of the consistency of this combination with the top-quark hypothesis is then evaluated using the \XWt variable:

\begin{equation}
\XWt = \sqrt{\left( \frac{m_W - 80~\GeV}{0.1 m_W} \right)^2 + \left( \frac{m_t - 173~\GeV}{0.1 m_t} \right)^2 },
\label{eqn:xwt}
\end{equation}
where $m_W$ is the invariant mass of the two-jet $W$ boson candidate and $m_t$ that of the three-jet top candidate.

All possible combinations of three jets are considered and the top-quark candidate with the smallest $X_{Wt}$ is chosen for each event. Events with the smallest $\XWt < 1.5$ are vetoed in the final selection. This requirement reduces the \ttbar contamination where both top quarks decay without leptons (\emph{hadronic}) by 60\%, and the \ttbar events that contain leptons (\emph{semileptonic}) by 45\%.

A correction is made using the known Higgs boson mass, where each Higgs boson candidate's four-momentum is multiplied by a correction factor $m_{H}/\mtwoj$. This leads to an improvement of approximately 30\% in signal \mfourj resolution with a significant reduction of low-mass tails caused by energy loss and with little impact on the background \mfourj shape.

The fraction of signal events accepted by the detector multiplied by the efficiency of each selection step is shown in Figure \ref{fig:resolvedEfficiency} for the narrow-width scalar, graviton, and SM non-resonant signal models. The acceptance times efficiency is higher for the graviton samples because spin-2 resonances decay more centrally, resulting in higher-\pt jets. The acceptance times efficiency is limited at low mass by the $p_{\mathrm{T}}$ requirement on the jets, and at high mass the chance to resolve four distinct jets becomes lower, and the $b$-tagging efficiency decreases.

\begin{figure*}
\begin{center}
\subfloat[SM non-resonant \HH]{
\includegraphics[width=70.00mm, trim = 0mm 0mm 0mm  0mm, clip]{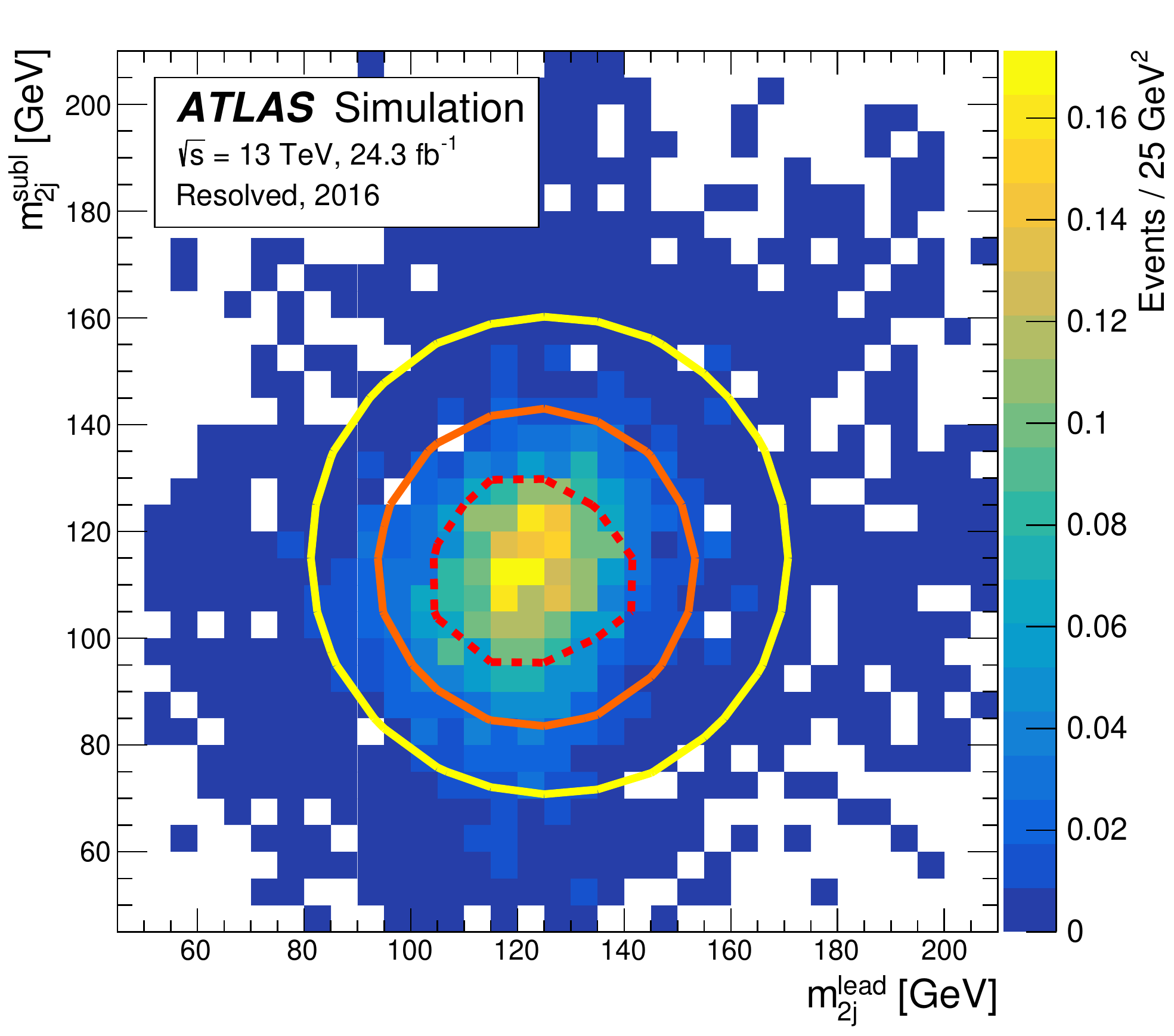}
}\hspace{5mm}
\subfloat[Multijet Background]{
\includegraphics[width=70.00mm, trim = 0mm 0mm 0mm  0mm, clip]{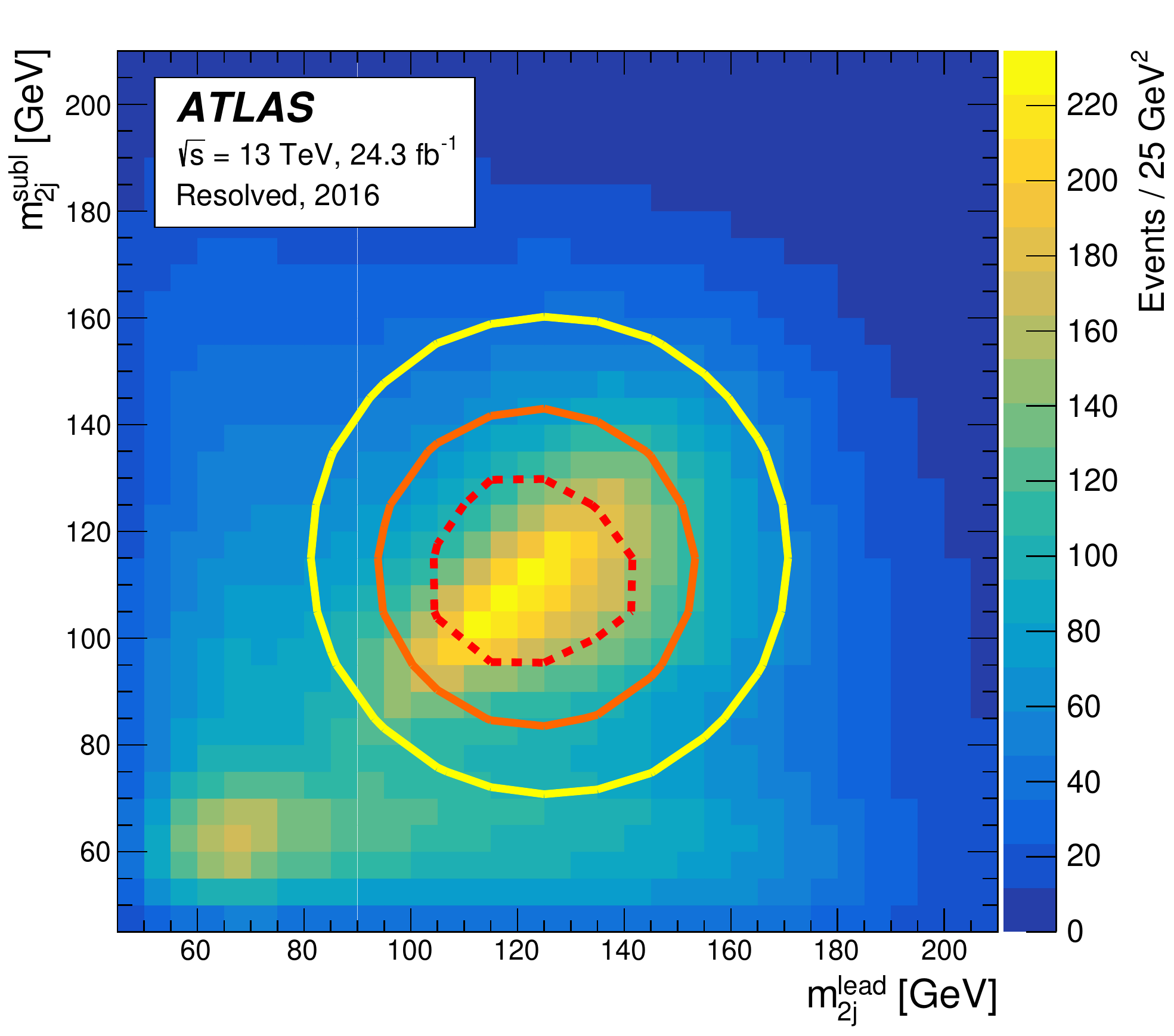}
}
\caption{Higgs boson candidate mass-plane regions. The signal region is inside the inner (red) dashed curve, the control region is outside the signal region and within the intermediate (orange) circle, the sideband is outside the control region and within the outer (yellow) circle.
(a) shows the SM non-resonant \HH\ process, and (b) shows the estimated multijet background, which is described in Section~\ref{sec:resolvedBKG}.}
\label{fig:resolvedRegions}
\end{center}
\end{figure*}

\begin{figure*}
\begin{center}
\subfloat[Scalar Signal]{\includegraphics[height=61.00mm, trim = 0mm 0mm 0mm  0mm, clip]{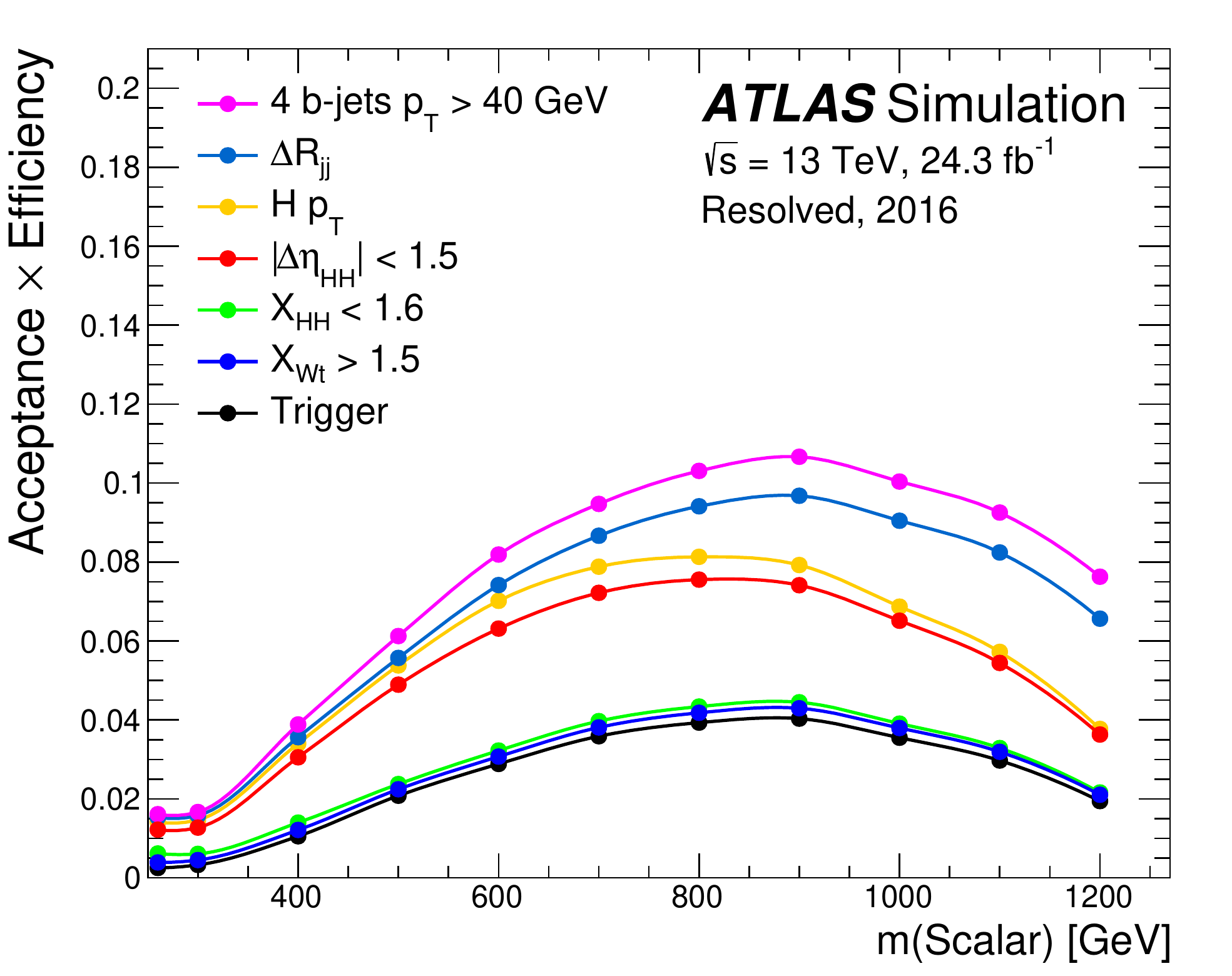}\hspace{0mm}}
\subfloat[Spin-2 Signal]{\includegraphics[height=61.00mm, trim = 12mm 0mm 0mm  0mm, clip]{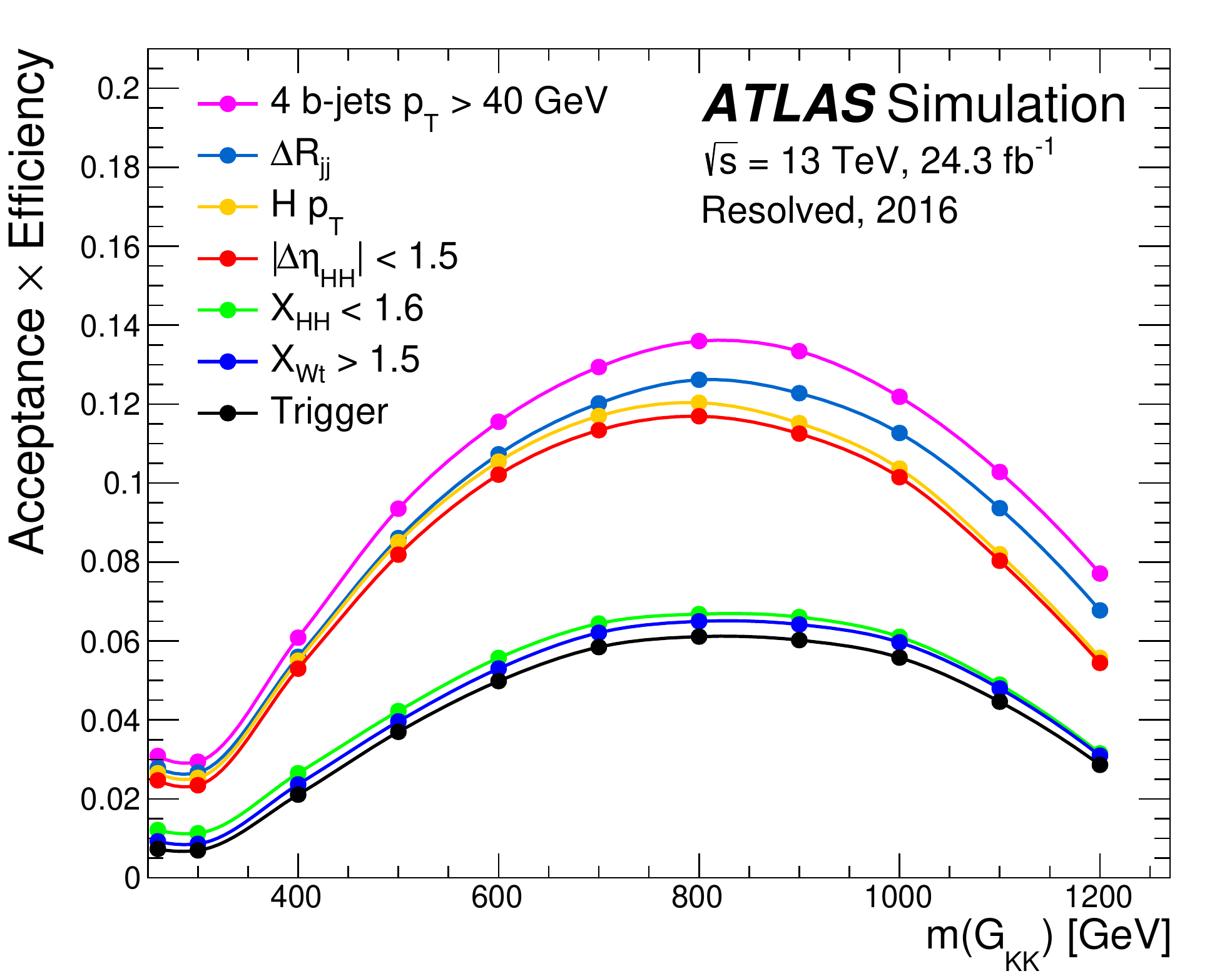}\hspace{-3mm}}
\subfloat[SM \HH]{\includegraphics[height=61.00mm, trim = 0mm 0mm 0mm  0mm, clip]{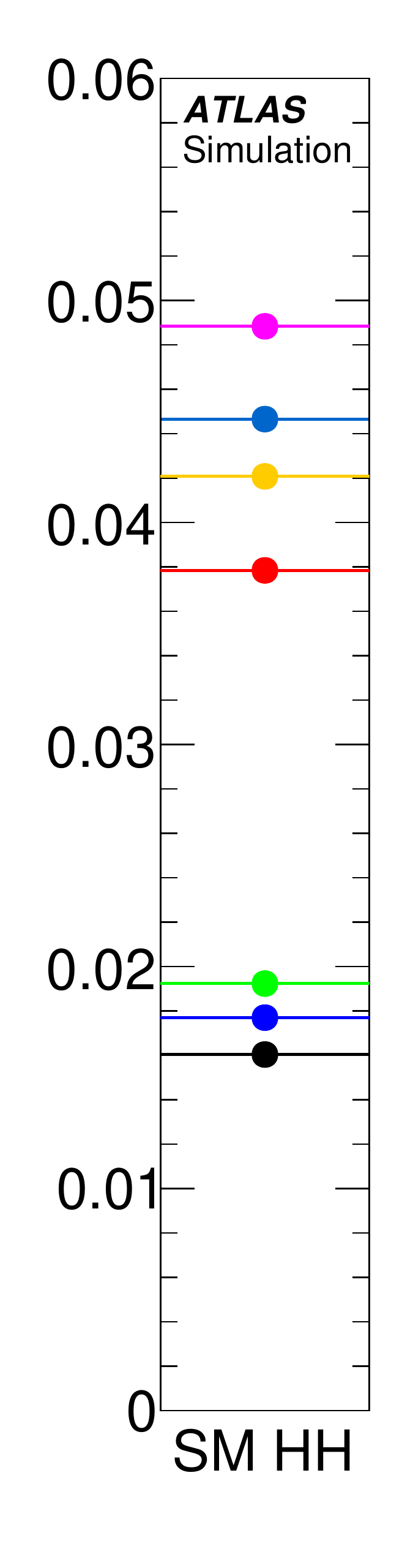}}
\caption{The selection acceptance times efficiency at each stage of the event selection for (a) a narrow-width scalar and (b) \Gtohhb with $\kMPl = 1$ for a range of resonance masses and (c) the SM non-resonant signal. Each selection step is detailed in Section~\ref{sec:resolvedSelection}.}
\label{fig:resolvedEfficiency}
\end{center}
\end{figure*}

\subsection{Background estimation}
\label{sec:resolvedBKG}
After the full event selection is applied, about $95\%$ of the background consists of multijet events, which are modelled using data. The remaining 5\% are \ttbar\ events. The \ttbar\ background normalization is determined from data, while the \mfourj spectrum is taken from simulation. A data-driven estimate of $Z+\mathrm{jets}$ events yields a contribution of 0.2\% to the total background, which is neglected in the following. Background from other sources, including processes involving single Higgs boson production, is also found to be negligible.

\subsubsection{Multijet background}
\label{sec:resolvedQCD}
The multijet background is modelled with an independent data sample selected using the same trigger and selection requirements as used in the signal region, except for the $b$-tagging requirement: at least four jets with $\pt > 40~\GeV$ are required, and exactly two of them have to be $b$-tagged.
This ``two-tag'' selection yields a data sample that consists of 88\% multijet events and 12\% \ttbar events.
The predicted signal contamination is negligible.

To model the multijet background in the four-tag sample with events selected in the two-tag sample, a product of two weights is applied to the two-tag events, where each weight corrects for different effects: additional jet activity and kinematic differences from applying $b$-tagging requirements.

The weights are derived in a signal-depleted sideband region of the \mlead--\msubl plane and validated using an orthogonal control region. The control region is defined as the region with
\begin{equation*}
\sqrt{(\mlead - 124~\GeV)^2 + (\msubl - 113~\GeV)^2}  < 30~\GeV,
\label{eqn:control}
\end{equation*}
excluding the signal region defined in Eq.~(\ref{eqn:resolvedXhh}). The sideband region is defined by
\begin{equation*}
\sqrt{(\mlead - 126~\GeV)^2 + (\msubl - 116~\GeV)^2}  < 45~\GeV,
\label{eqn:sideband}
\end{equation*}
excluding the control and signal regions. The outer boundaries of the regions are selected to achieve sufficient statistical precision while ensuring that the kinematic properties of the sideband region are representative of the signal region. The shifted centres of these regions [c.f. (120~\GeV, 110~\GeV) for the signal region] ensure the mean Higgs boson candidate masses are equal in the three regions.

The event weight that corrects for additional jet activity is obtained as follows. 
For each event, all possible combinations are considered where at least two anti-$b$-tagged jets, that pass the kinematics requirements, are treated as $b$-jets. A constant per-jet transfer factor, $f$, is assigned to each of these jets and a factor (1-$f$) to the remaining ones, and a weight for the event is computed from the sum of all combinations. One of the combinations is then randomly chosen to form the Higgs boson candidates using the same procedure as in the four-tag sample. The transfer factor is determined by comparing the jet multiplicity distributions of the two-tag and four-tag selections in the sideband region. The resulting value is reported in Table~\ref{tab:resolvedNORMFIT}.

The events in the two-tag data sample are then weighted further to correct for the kinematic differences caused by the additional $b$-tagging requirement of the four-tag sample. These differences can arise for the following reasons: the $b$-tagging efficiency varies as a function of jet \pt and $\eta$; the various multijet processes contribute with different fractions in each sample; and the fractions of events accepted by each trigger path changes. 

The weights are determined by fitting cubic splines to the ratio of kinematic distributions of the total background model to data after subtracting the \ttbar\ contribution in both samples (12\% in the two-tag sideband and 7\% in the four-tag sideband), before the cut on \XWt. This is done for five distributions that show large differences between the four-tag and two-tag samples. These are: the average $|\eta|$ of the four jets constituting the Higgs boson candidate; the \pt of the second and fourth leading (in \pt) constituent jets; the smallest $\Delta R$ between any two constituent jets; and the $\Delta R$ between the other two constituent jets. The reweighting is performed using one-dimensional distributions, and is iterated until the weights converged to stable values.

Agreement of the background model and data after these reweightings is checked in the control and sideband regions, and is found to be notably improved. This improvement is verified in the variables used to derive the weights and additionally in many other kinematic distributions.

\subsubsection{Background normalization and the \ttbar background}
\label{sec:resolvedTTbar}

The \mfourj distribution of the \ttbar background is modelled using simulation. To improve the statistical precision, the two-tag simulated distribution is used for the hadronic \ttbar sample, after correction by the same kinematic weights used for the multijet model. This procedure is validated in an inclusive region that contains events from signal region, sideband and control region; and good agreement is observed between the corrected two-tag sample and the four-tag sample in several distributions. The semileptonic \ttbar background is modelled directly, using the four-tag MC sample of the \ttbar\ background.

The normalizations of the \ttbar and multijet backgrounds are determined simultaneously by fitting the yields of semileptonic \ttbar, hadronic \ttbar, and multijet events in three background-enriched samples. These background-enriched samples are all defined as having Higgs boson candidate's masses in the sideband region, but consistent results are found using the control region data. Specific background-enriched samples are defined with requirements additional to the sideband selection. The semileptonic enriched sample must contain an isolated muon \cite{PERF-2015-10} with $\pt > 25~\GeV$ and from Eq.~(\ref{eqn:xwt}), $\XWt < 1.5$. The sample enriched in hadronic \ttbar requires $\XWt < 0.75$, while the sample defined by $\XWt > 0.75$ is enriched in multijet events.

There are three parameters in the normalization fit: \muqcd, which scales the multijet yield from the two-tag to the four-tag sideband region after the per-jet transfer factor $f$ and the kinematic weights have been applied; and two parameters $\alpha_{t\bar{t}}^{\mathrm{hadronic}}$ and $\alpha_{t\bar{t}}^{\mathrm{semileptonic}}$ that correct the normalizations of the yields for the hadronic and semileptonic \ttbar MC samples in the four-tag sideband region.

The result of the normalization fit is presented in Table~\ref{tab:resolvedNORMFIT}. The uncertainties are those from limited data and MC sample sizes, and they are propagated to the final fit, as described in Section~\ref{sec:statAna}.

\begin{center}
\begin{table}[htbp!]
\caption{The fitted values of the normalization parameters \muqcd and $\alpha$ of both \ttbar\ samples and their statistical uncertainties, given for the two datasets. The per-jet transfer factor $f$ is also listed, and is explained in Section~\ref{sec:resolvedQCD}. }
\begin{center}
\begin{tabular}{lcccc}
\toprule
Dataset & $f$ & \muqcd & $\alpha_{t\bar{t}}^{\mathrm{hadronic}}$ & $\alpha_{t\bar{t}}^{\mathrm{semileptonic}}$\\
\midrule
2015 & 0.22 & $0.0838\pm0.0038$ & $1.19\pm0.45$ & $1.44\pm0.48$ \\
2016 & 0.15 & $0.2007\pm0.0031$ & $1.15\pm0.25$ & $1.7\pm0.19$ \\
\bottomrule
\end{tabular} 
\end{center}
\label{tab:resolvedNORMFIT}
\end{table}
\end{center}

After the reweighting corrections and the application of the normalization, there is good agreement between the background model and the data distributions in the sideband as well as the control regions. The distributions of \mfourj\ in the control region for both datasets are displayed in Figure~\ref{fig:resolvedCRm4j}.

\begin{figure}[!ht]
\begin{center}
\subfloat[2015 dataset]{\includegraphics[width=0.5\textwidth]{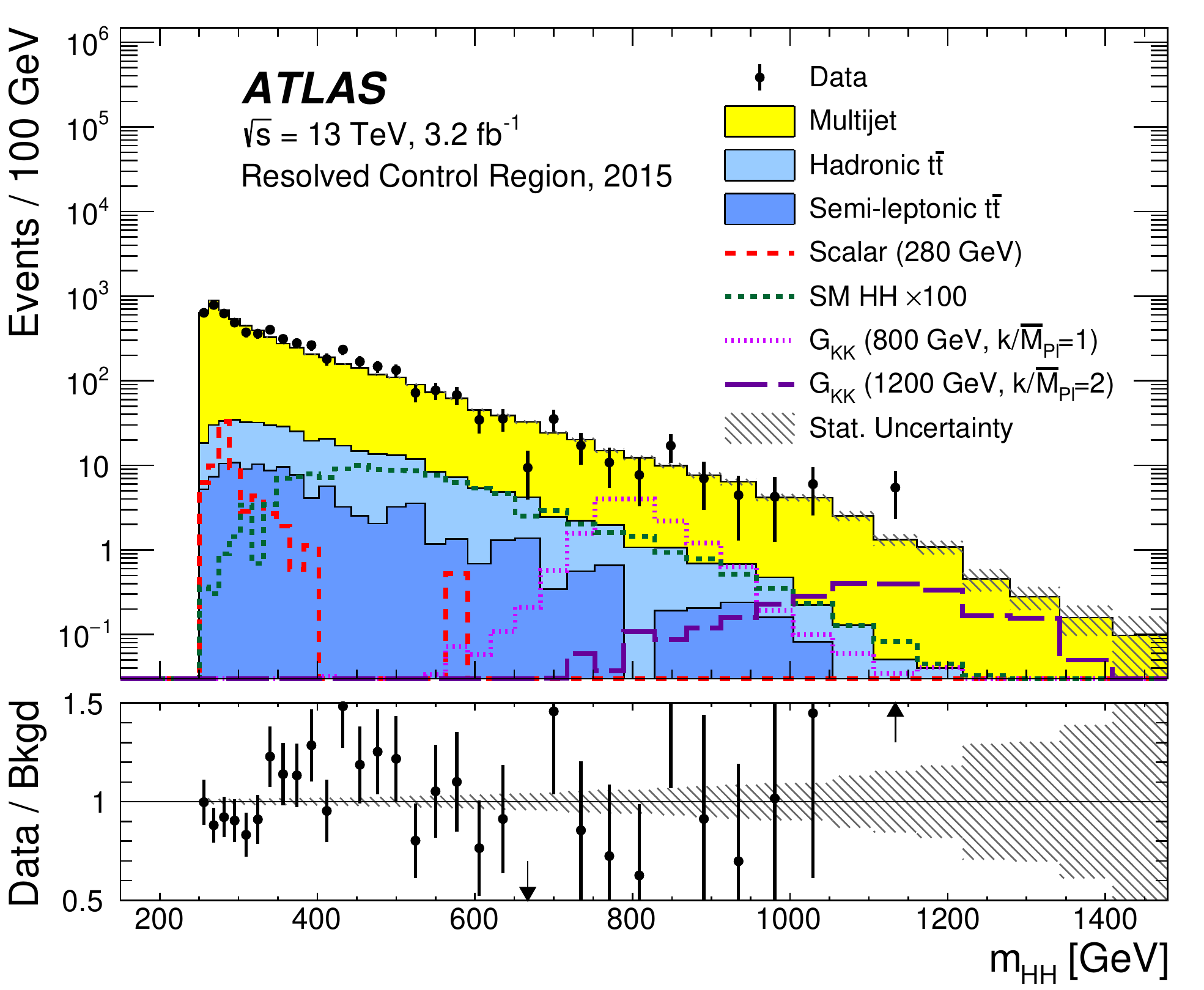}}
\subfloat[2016 dataset]{\includegraphics[width=0.5\textwidth]{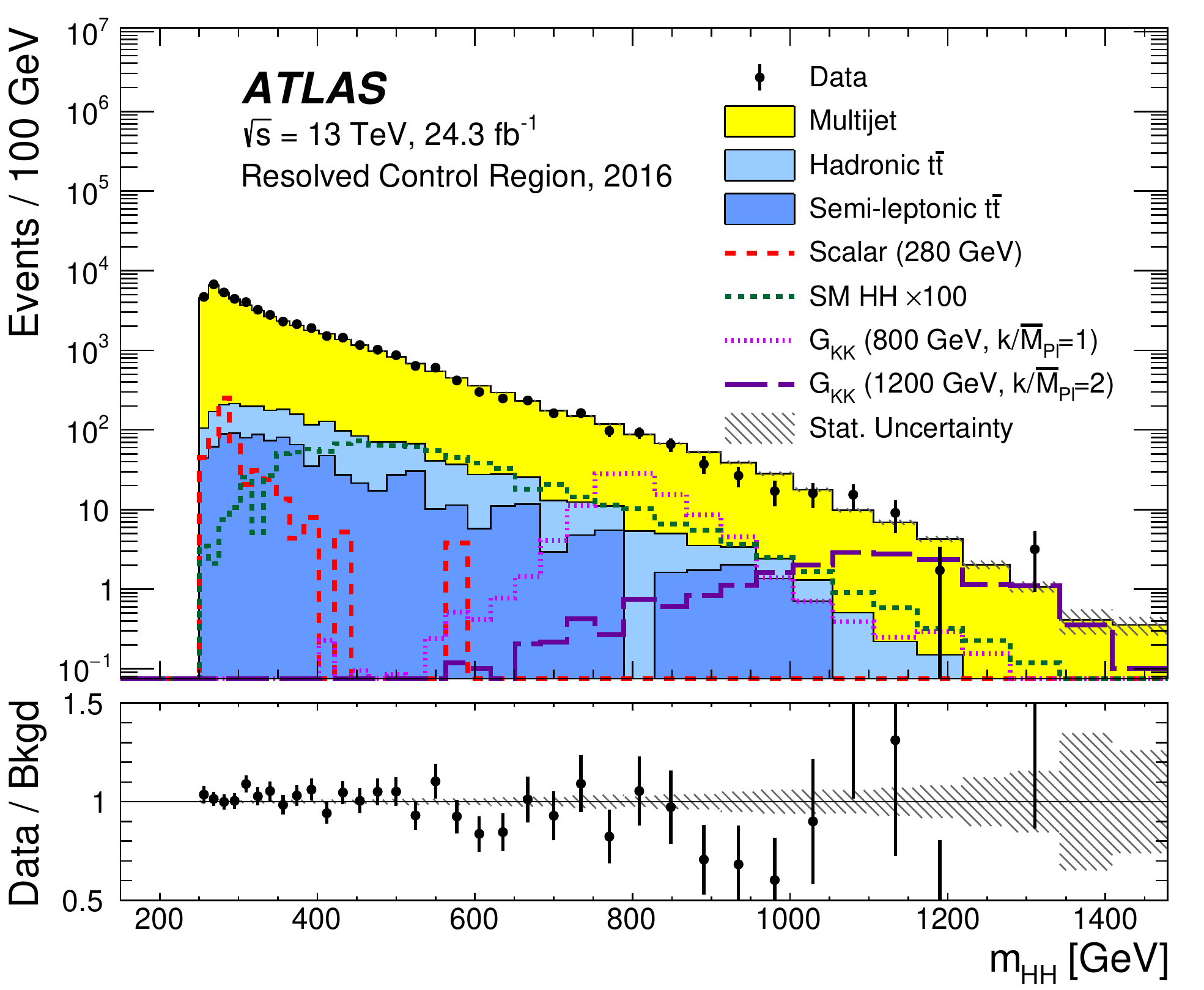}}
\caption{Distributions of \mfourj in the control region of the resolved analysis for (a) 2015 data and (b) 2016 data, compared to the predicted backgrounds. The hatched bands represent the statistical uncertainties. The expected signal distributions of \Grav\ resonances with masses of 800 and 1200~\GeV, a 280~\GeV\ scalar particle and SM non-resonant \HH\ production $(\times 100)$ are also shown. The scalar sample is normalized to a cross section times branching ratio of 2.7~pb.}
\label{fig:resolvedCRm4j}
\end{center}
\end{figure}

\subsection{Systematic uncertainties}
\label{sec:resolvedSyst}

Background uncertainties are propagated from the fit which determines the multijet and \ttbar yields. The statistical uncertainties in the scale factors (Table~\ref{tab:resolvedNORMFIT}) are propagated including the correlations, by calculating three orthogonal eigenvariations from the covariance matrix of the normalization fit, resulting in three nuisance parameters, such that each parameter acts on the three background normalizations simultaneously.

Shape uncertainties in the multijet background are assessed by deriving an alternative background model using the same procedure as in the nominal case, but using data from the control region rather than from the sideband. This alternative model and the baseline are consistent with the observed data in their regions and with each other. The differences between the baseline and the alternative are used as a background-model shape uncertainty, with a two-sided uncertainty defined by symmetrizing the difference about the baseline. The uncertainty is split into two components to allow two independent variations: a low-\Ht and a high-\Ht component, where \Ht is the scalar sum of the \pt of the four jets constituting the Higgs boson candidates. The boundary value is 300~\GeV. The low-\Ht shape uncertainty primarily affects the \mfourj spectrum below 400~\GeV\ (close to the kinematic threshold) by up to 5\%, and the high-\Ht uncertainty mainly \mfourj above this by up to 30\% relative to nominal. 

Shape uncertainties affecting the \ttbar background component are dominated by those associated with the use of two-tag simulation to model the \mfourj distribution of hadronic \ttbar. Since this background is also reweighted, its uncertainties can be assessed using the same procedure as for the multijet background and are again split into low-\Ht and high-\Ht components. The impact of detector and theoretical modelling uncertainties on the \ttbar background shape were assessed but are found to be negligible, because the data-driven reweighting procedure used for the multijet modelling compensates for biases in the \ttbar sample by adjusting the multijet model.

Theoretical uncertainties in the signal acceptance result from variations of renormalization and factorization scales, PDF set uncertainties, and uncertainties in modelling of the underlying event and hadronic showers (therefore varying initial- and final-state radiation). The scales are varied by factors of 2 and 0.5. The PDF uncertainties are evaluated using \textsc{PDF4LHC15} sets~\cite{pdfs}. The parton shower and underlying event are varied by replacing \textsc{Herwig++} with \PYTHIA for the scalar samples, and vice versa for the spin-2 samples. The total theoretical uncertainty is dominated by the shower variations. The size of the variation of the expected signal yield is typically below 10\% but can increase to 23\%, depending on the signal hypothesis. 

The following detector modelling uncertainties are evaluated: uncertainties in the jet energy scale (JES) and resolution (JER), uncertainties in the $b$-tagging efficiency, and uncertainties in the trigger efficiency. The jet energy uncertainties are derived using in situ measurement techniques described in Refs. \cite{ATLAS-CONF-2015-057, ATLAS-CONF-2015-017, ATLAS-CONF-2015-037}. The JES systematic uncertainty is evaluated following the prescription outlined in Ref.~\cite{Aad:2014bia}. The JER uncertainty is evaluated by smearing jet energies according to the systematic uncertainties of the resolution measurement~\cite{Aad:2014bia}.

The uncertainty in the integrated luminosity for the 2015 (2016) dataset is $\pm2.1$\% ($\pm2.2$\%), which was evaluated using a technique similar to that described in Ref.~\cite{LumiCiteUP}.

The uncertainty in the $b$-tagging efficiency is evaluated by propagating the systematic uncertainty in the \pt-dependent, measured tagging efficiency for $b$-jets~\cite{ATLAS-CONF-2014-004}. For $b$-jets with \pt~$> 300$~\GeV, systematic uncertainties in the tagging efficiencies are extrapolated with simulation and are consequently larger~\cite{Aad:2015ydr}. Uncertainties arising from mis-tagging jets that do not contain $b$-hadrons are negligible.

Trigger efficiency uncertainties are evaluated for the signal, based on the systematic uncertainties arising from the per-jet online \btagging\ measurements. There is an additional, small, non-closure uncertainty associated with the calculation of per-event trigger efficiencies using the measured per-jet efficiencies. The total trigger efficiency uncertainty is $\pm2\%$ for the non-resonant signal and for resonant signals with masses below 1100~\GeV, growing to $\pm5\%$ for a resonance of mass 1400~\GeV.

Uncertainties in the signal are fully correlated between the 2015 and 2016 datasets, except those for the luminosity and trigger efficiency. Systematic uncertainties in the normalization and shape of the multijet and \ttbar background models are treated as uncorrelated between the two datasets. The case of an unknown, partial correlation can be neglected because the sensitivity of the 2015 dataset is much smaller than that of 2016.

Table~\ref{tab:resolvedSyst} summarizes the relative impact of the uncertainties on the event yields.
\begin{table}[ht!]
\scriptsize
\begin{center}
\caption{Summary of systematic relative uncertainties (expressed in percentage yield) in the total background and signal event yields in the signal region of the resolved analysis. Uncertainties are provided for both the 2015 and 2016 analyses for background, a \Grav resonance with \kMPl = 1 and $m(G_{\mathrm{KK}})=800$~\GeV, a scalar with a mass of 280~\GeV\ and SM non-resonant Higgs boson pair production. The total uncertainties include the effect of correlations.}

\resizebox{\textwidth}{!}
{
\begin{tabular}{l*8{S[table-format=2.1, table-number-alignment=center, round-mode=places, round-precision=1]}}
\toprule
      & \multicolumn{4}{c}{2015} & \multicolumn{4}{c}{2016} \\
Source      & {Background} & {Scalar} & {SM \HH} & {\Grav} & {Background} & {Scalar} & {SM \HH} & {\Grav}  \\
\midrule
Luminosity    &      {--}   &  2.1          &  2.1       &    2.1        &   {--}       &  2.2       & 2.2       &  2.2  \\
Jet energy    &      {--}   & 17            &  7.1       &    3.7        &   {--}       & 17         & 6.4       &  3.7  \\
$b$-tagging   &      {--}   & 13            & 12         &   14          &   {--}       & 13         & 12        & 14    \\
$b$-trigger   &      {--}   &  4.0          &  2.3       &    1.3        &   {--}       &  2.6       &  2.5      &  2.5  \\
Theoretical   &      {--}   & 23            &  7.2       &    0.6        &   {--}       & 23         &  7.2      &  0.6  \\
Multijet stat &      4.2    &  {--}         &  {--}      &    {--}       &   1.5        &  {--}      &  {--}     &  {--} \\
Multijet syst  &      6.1    &  {--}         &  {--}      &    {--}       &   1.8        &  {--}      &  {--}     &  {--} \\
\ttbar stat   &      2.1    &  {--}         &  {--}      &    {--}       &   0.78       &  {--}      &  {--}     &  {--} \\
\ttbar syst    &      3.5    &  {--}         &  {--}      &    {--}       &   0.26       &  {--}      &  {--}     &  {--} \\
\midrule
Total         &      7.5    & 31            &  16        &   15          &   1.8        & 31         &  16       &  15   \\

\bottomrule
\end{tabular}
}
\label{tab:resolvedSyst}
\end{center}
\end{table}
\subsection{Signal region event yields}
The number of events observed in the data, the predicted number of background events in the signal region, and the predicted yield for three potential signals are presented in Table~\ref{tab:resolvedResults} for both the 2015 and 2016 datasets. The numbers of observed and predicted events in the control region are also given, and they are in agreement. A discrepancy between data and the total prediction is seen in the 2016 dataset; about half of this excess can be attributed to one bin at $\mfourj=280$~\GeV.

\begin{table}[!ht]
\begin{center}
\caption{The number of predicted background events in the signal region (SR) for the resolved analysis compared to the data, for the 2015 and 2016 datasets. The yields for three potential signals, an 800~\GeV\ \Grav resonance with \kMPl = 1, a scalar with a mass of 280~\GeV, and SM non-resonant Higgs boson pair production, are also shown. The scalar sample is normalized to a cross section times branching ratio of 2.7~pb. The quoted uncertainties include both the statistical and systematic uncertainties, and the total uncertainty considers correlations. The numbers of observed and predicted events are also given in the control region (CR).}

\begin{tabular}{lS[table-format=3.3,table-number-alignment=center]@{\,$\pm\,$}S[table-format=2.3, table-number-alignment=left]
S[table-format=4.2,table-number-alignment=center]@{\,$\pm\,$}S[table-format=3.2, table-number-alignment=left]
S[table-format=3.0,table-number-alignment=center]@{\,$\pm\,$}S[table-format=3.0, table-number-alignment=left]
S[table-format=4.0,table-number-alignment=center]@{\,$\pm\,$}S[table-format=3.0, table-number-alignment=left]}
\toprule

Sample & \multicolumn{2}{c}{2015 SR} & \multicolumn{2}{c}{2016 SR} & \multicolumn{2}{c}{2015 CR} & \multicolumn{2}{c}{2016 CR} \\

\midrule

Multijet                & 866     &  70      &  6750 & 170  & 880 & 71 & 7110 & 180 \\
\ttbar, hadronic        &  52     &  35      & 259   & 57   & 56  & 37 & 276  & 61 \\
\ttbar, semileptonic    &  13.9   &  6.5     &  123  &  30  & 20  & 9  & 168 & 40 \\
\midrule

Total         & 930 & 70      & 7130 & 130  & 956 & 50 &  7550 & 130 \\

\midrule

Data         & \multicolumn{2}{l}{928}    & \multicolumn{2}{l}{7430} & \multicolumn{2}{l}{969} &\multicolumn{2}{l}{7656}  \\

\midrule

\Grav$\left(800~\GeV\right)$ & 12.5   & 1.9     &  89  & 14 \\
Scalar (280~\GeV)            & 24.0   & 7.5     & 180  & 57 \\
SM \HH                       & 0.607  & 0.091   & 4.43 & 0.66 \\

\bottomrule
\end{tabular}
\label{tab:resolvedResults}
\end{center}
\end{table}

Figure~\ref{fig:resolvedHHUnblinded} shows comparisons of the predicted \mfourj~background distributions to those observed in the 2015 and 2016 datasets. A few signal models are also displayed. The scalar sample shown is normalized to a cross section times \hbb\ branching ratio of 2.7~pb, which is the best-fit value (the fit is described in Section~\ref{sec:statAna}). The predicted background and observed distributions are mostly in agreement.

\begin{figure}[!ht]
\begin{center}
\subfloat[2015 dataset]{\includegraphics[width=0.5\textwidth]{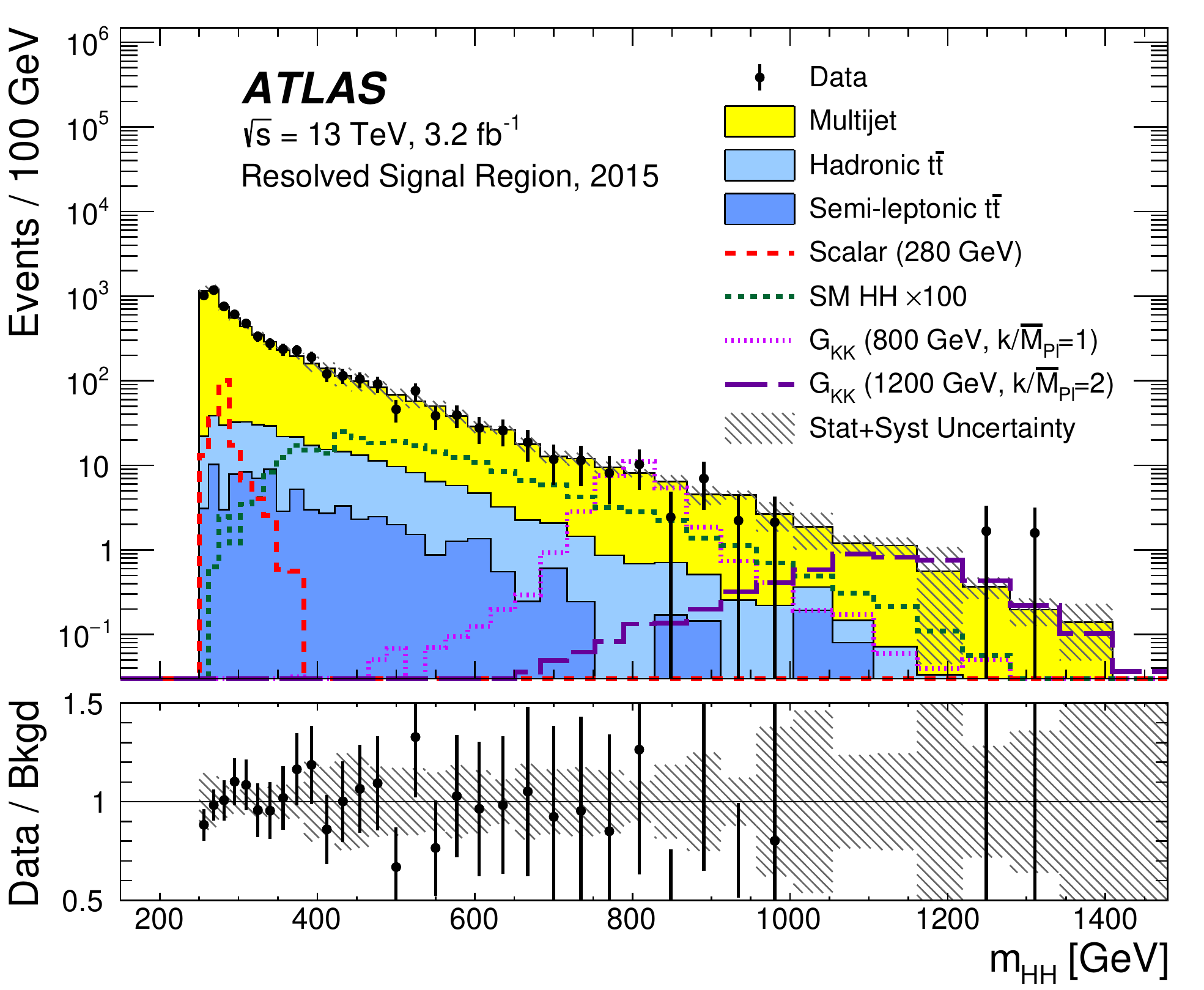}}
\subfloat[2016 dataset]{\includegraphics[width=0.5\textwidth]{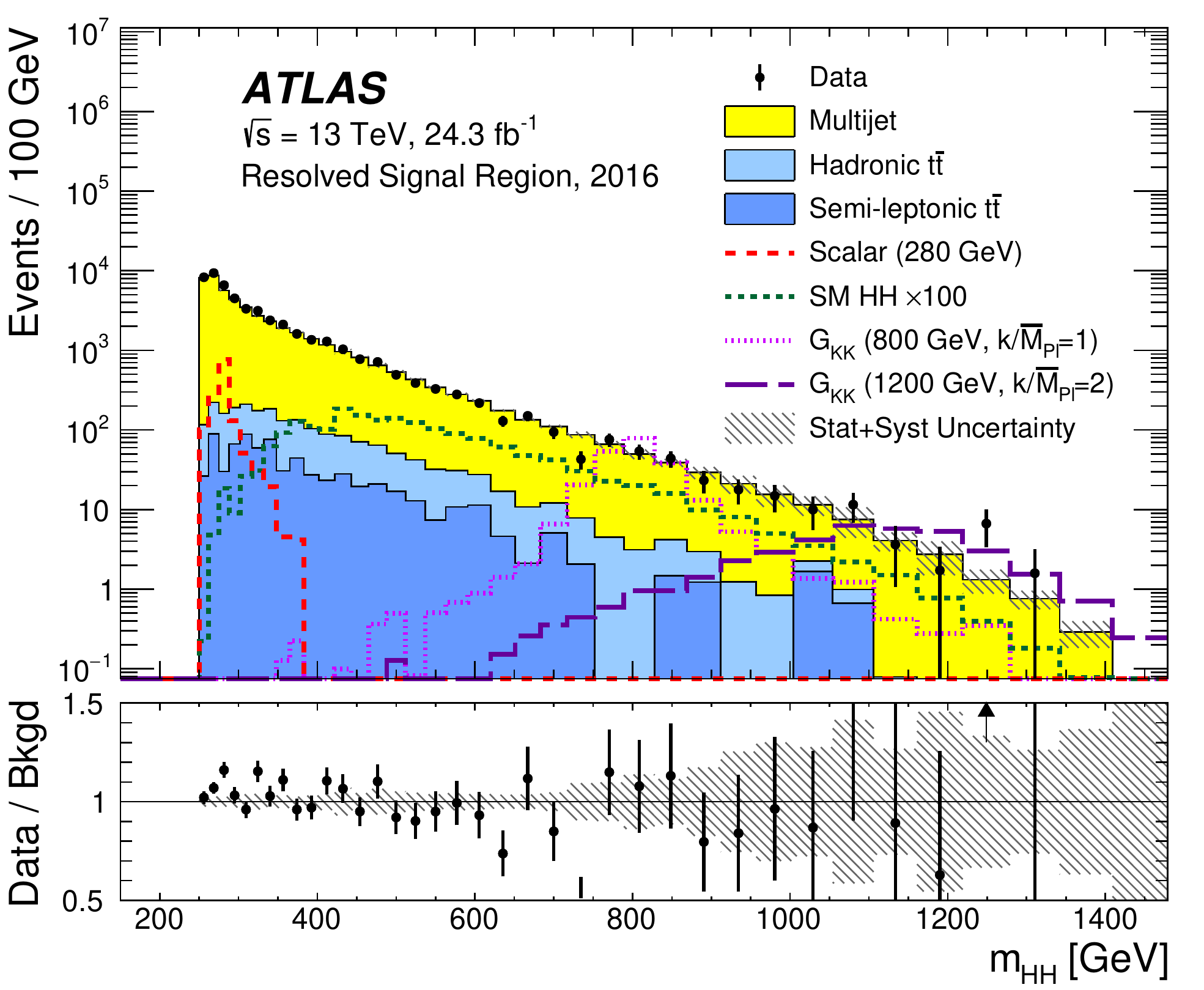}}
\caption{Distributions of \mfourj in the signal region of the resolved analysis for (a) 2015 data and (b) 2016 data, compared to the predicted backgrounds. The hatched bands represent the combined statistical and systematic uncertainties in the total background estimates. The expected signal distributions of \Grav\ resonances with masses of 800 and 1200~\GeV, a 280~\GeV\ scalar sample and SM non-resonant \HH\ production $(\times 100)$ are also shown. The scalar sample is normalized to a cross section times branching ratio of 2.7~pb.}
\label{fig:resolvedHHUnblinded}
\end{center}
\end{figure}

\FloatBarrier
\section{Boosted analysis}
\label{sec:boosted}

The boosted analysis is optimized to discover signals arising from production of high-mass resonances decaying into Higgs boson pairs. The strategy is to select two Higgs boson candidates with mass near \mH, each composed of a single large-\R jet with at least one $b$-tagged track-jet matched to it. Three samples are defined according to the total number of $b$-tagged track-jets associated with the Higgs boson candidates. Since the triggers are fully efficient for the signal processes in both the 2015 and the 2016 datasets, the two datasets are combined into one.

The invariant mass of the two-Higgs-boson-candidate system, \mtwoJ, is used as the final discriminant between Higgs boson pair production and the SM backgrounds. Events that pass the resolved signal region selection are vetoed in the boosted analysis, thus priority is given to the resolved analysis if an event passes both selections, which increases the sensitivity.

\subsection{Selection}

Events are required to have a primary vertex with at least two tracks matched to it. Events are selected that have at least two \akt large-\R jets with $ \pt > 250~\GeV$, $|\eta| <$ 2.0, and mass $\mJ > 50~\GeV$. Only the two jets with highest \pt are retained for further selection. The leading jet is required to have $\pt > 450~\GeV$, which ensures 100\% trigger efficiency. Since high-mass resonances tend to produce jets that are more central than those from multijet background processes, the two large-$R$ jets are required to have a separation $|\Delta \eta| <$ 1.7. To be considered as a Higgs boson candidate, each large-$R$ jet must contain at least one $b$-tagged $R=0.2$ track-jet matched to it by ghost association. 

Three separate samples of events are selected. The ``two-tag'' sample requires each Higgs boson candidate to have exactly one associated $b$-tagged track-jet. The ``three-tag'' and ``four-tag'' samples require that there are exactly three or exactly four $b$-tagged track-jets associated with Higgs boson candidates in the event, with two $b$-tagged track-jets associated with one candidate and one or two associated with the other candidate. Increasing the required number of associated $b$-tagged track-jets in the event increases signal purity at the expense of lower signal efficiency. This loss of efficiency is particularly pronounced for the highest resonance masses. It is rare to identify four distinct track-jets containing $b$-hadrons in these high-mass events due to the inefficiency in $b$-tagging high-\pt track-jets, and also because the extremely high Lorentz boosts make it difficult to resolve each $b$-quark as a separate track-jet.

Signal event candidates are selected if each of the large-$R$ jets has a mass consistent with that of the Higgs boson. This is defined analogously to the resolved analysis in Eq.~(\ref{eqn:resolvedXhh}):
\begin{equation*}
\label{eq:boostedxhh}
\Xhh = \sqrt{\left(\frac{\mleadJ - 124~\GeV}{0.1\mleadJ}\right)^2 + \left(\frac{\msublJ - 115~\GeV}{0.1\msublJ}\right)^2} < 1.6,
\end{equation*}
with the large-\R jet masses \mleadJ and \msublJ, where leading (sub-leading) refers to the large-\R jet with the largest (second largest) \pT. The mass resolution of the Higgs boson candidates found in simulation is similar to that in the resolved analysis. The central values of 124~\GeV\ and 115~\GeV\ correspond to the median values of the narrowest intervals that contain 90\% of the simulated signal. The requirement of $X_{\HH} <$ 1.6 is chosen such that it optimizes the sensitivity, and it defines the signal region in the leading--sub-leading Higgs boson candidate mass plane.

The fraction of signal events accepted by the detector multiplied by the efficiency of each selection step for the $G_{\mathrm{KK}}$ model with $\kMPl = 1$ is shown in Figure~\ref{fig:boostedEfficiency}, and the efficiency for signal events to populate samples with either two, three or four $b$-tagged track jets is displayed in Figure~\ref{fig:boostedEfficiency_nTags}.

A correction is made, multiplying the four-momentum of each large-\R jet by a factor $m_{H}/m_{\mathrm{J}}$. This slightly improves the resolution of \mtwoJ for signal events by reducing the low-mass tails caused by energy loss. There is little impact on the background distribution. 

\begin{figure*}
\begin{center}
\subfloat[]{\includegraphics[width=0.48\textwidth]{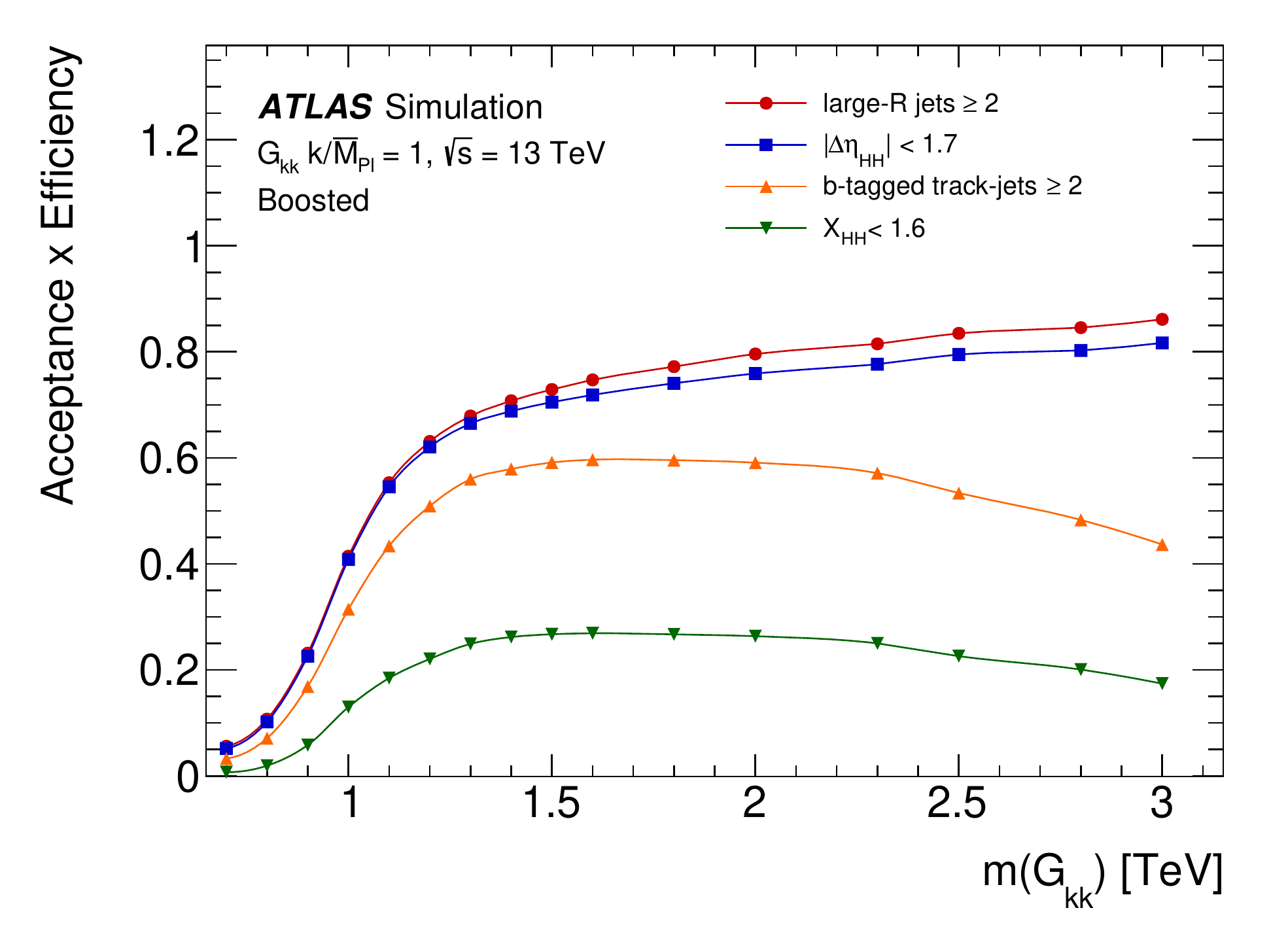}\label{fig:boostedEfficiency}}
\subfloat[]{\includegraphics[width=0.48\textwidth]{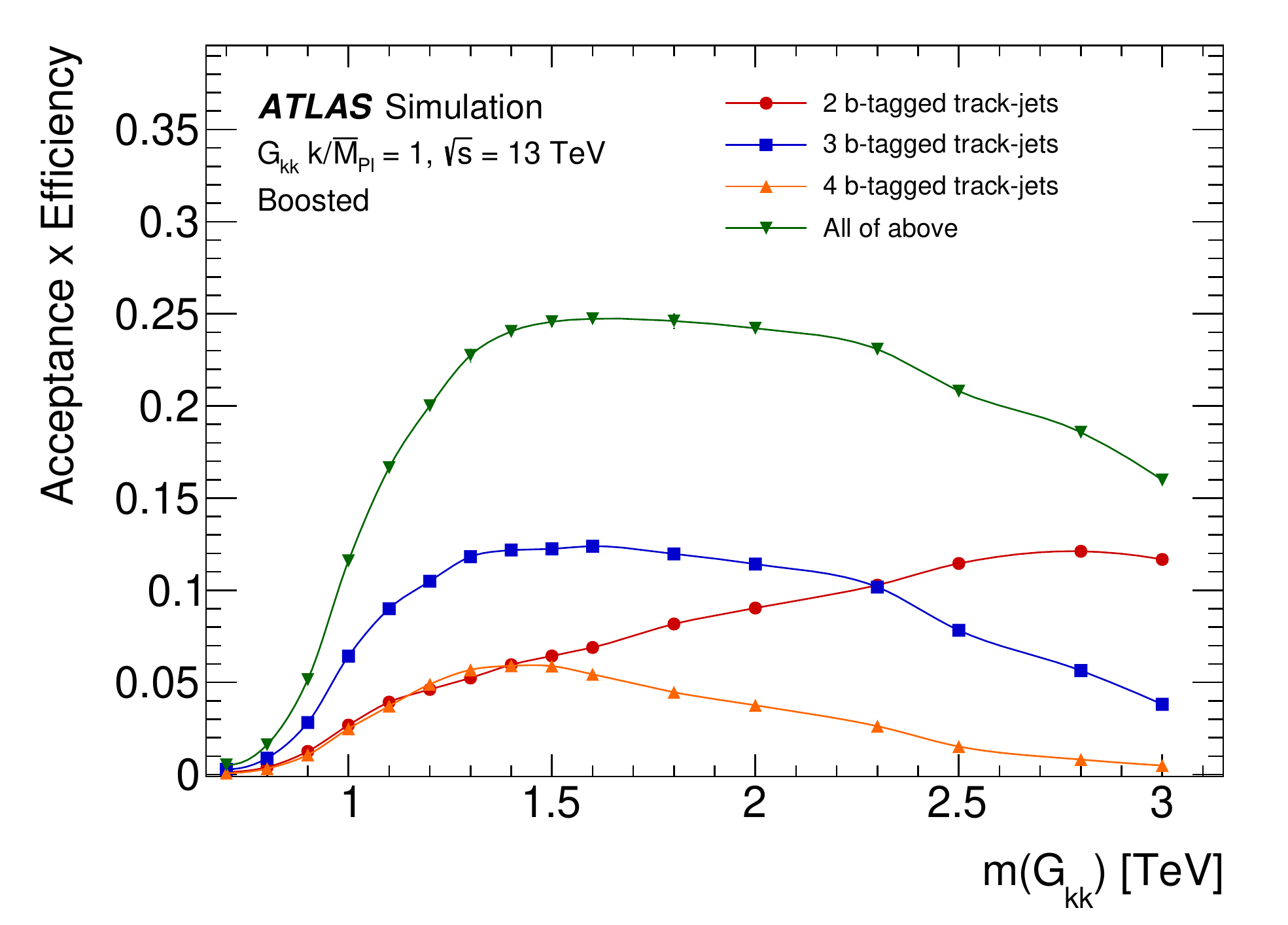}\label{fig:boostedEfficiency_nTags}}
\caption{(a) The selection acceptance times efficiency of the boosted analysis at each stage of the event selection as a function of the generated graviton mass for $\kMPl = 1$. The trigger efficiency is approximately 100\% after the requirement of two large-$R$ jets, so it is not shown. 
(b) The selection efficiency when requiring two, three or four $b$-tagged track jets associated to the two large-\R jets, as a function of the generated graviton mass for $\kMPl = 1$. In both figures for the case of two $b$-tagged track jets, they can either both be associated to the same, or one to each large-\R jet.
}
\end{center}
\end{figure*}

\subsection{Background estimation}
\label{sec:boostedBackground}
The main backgrounds after selection are multijet events, which comprise 80--95\% of the total background, depending on the number of $b$-tagged track-jets required, with the remainder being \ttbar events. The contribution of $Z+\mathrm{jets}$ events to the signal region in each sample is estimated using simulation to be less than 0.1\%, and is therefore neglected. Other sources of background, including processes involving single Higgs boson production, are also negligible. The multijet events are modelled using data. The \ttbar yield is estimated using a data-driven technique, while the \ttbar \mtwoJ shape is taken from simulation. 

The shape of the multijet background is modelled using independent data samples selected with the same trigger and selection requirements as described above, but with fewer $b$-tagged track-jets. To model the two-tag sample, a ``1$b$-1'' sample is selected comprising events where one of the large-$R$ jets contains a single $b$-tagged track-jet, while the other large-$R$ jet contains no $b$-tagged track-jets, but at least one track-jet which fails $b$-tagging. Analogous ``2$b$-1'' and ``2$b$-2'' samples are selected to model the three- and four-tag samples, respectively. These comprise events where one large-$R$ jet contains two $b$-tagged track-jets and the other large-$R$ jet contains no $b$-tagged track-jets but exactly one (2$b$-1) or at least two (2$b$-2) track-jets that fail $b$-tagging. These selections are referred to below as lower-tagged samples.

The normalizations of both the multijet and the \ttbar backgrounds are derived using a signal-free sideband region. The definition of the sideband region is optimized such that it contains the bulk of the \ttbar\ events, yet is close enough to the signal region to accurately model the background kinematics there. A control region between the signal and sideband regions is used to validate the background models and to assign systematic uncertainties. The regions are defined using \Xhh and two other variables:

\begin{equation*}
\label{eq:boostedrhh}
R_{\HH} = \sqrt{ (\mleadJ - 124~\GeV)^2 + (\msublJ - 115~\GeV)^2 }, \quad
R_{\HH}^{\text{high}} = \sqrt{ (\mleadJ - 134~\GeV)^2 + (\msublJ - 125~\GeV)^2 }.
\end{equation*}

The signal region contains events with $X_{\HH}$ < 1.6, events in the control region fulfil $R_{\HH}$ < 33~\GeV and $X_{\HH}$ > 1.6, and events in the sideband satisfy $R_{\HH}$ > 33~\GeV and $R_{\HH}^{\text{high}}$ < 58~\GeV. The three regions of the \mleadJ--\msublJ plane are depicted in Figure \ref{fig:boostedRegions}.

\begin{figure}[!ht]
\begin{center}
\includegraphics[width=0.60\textwidth]{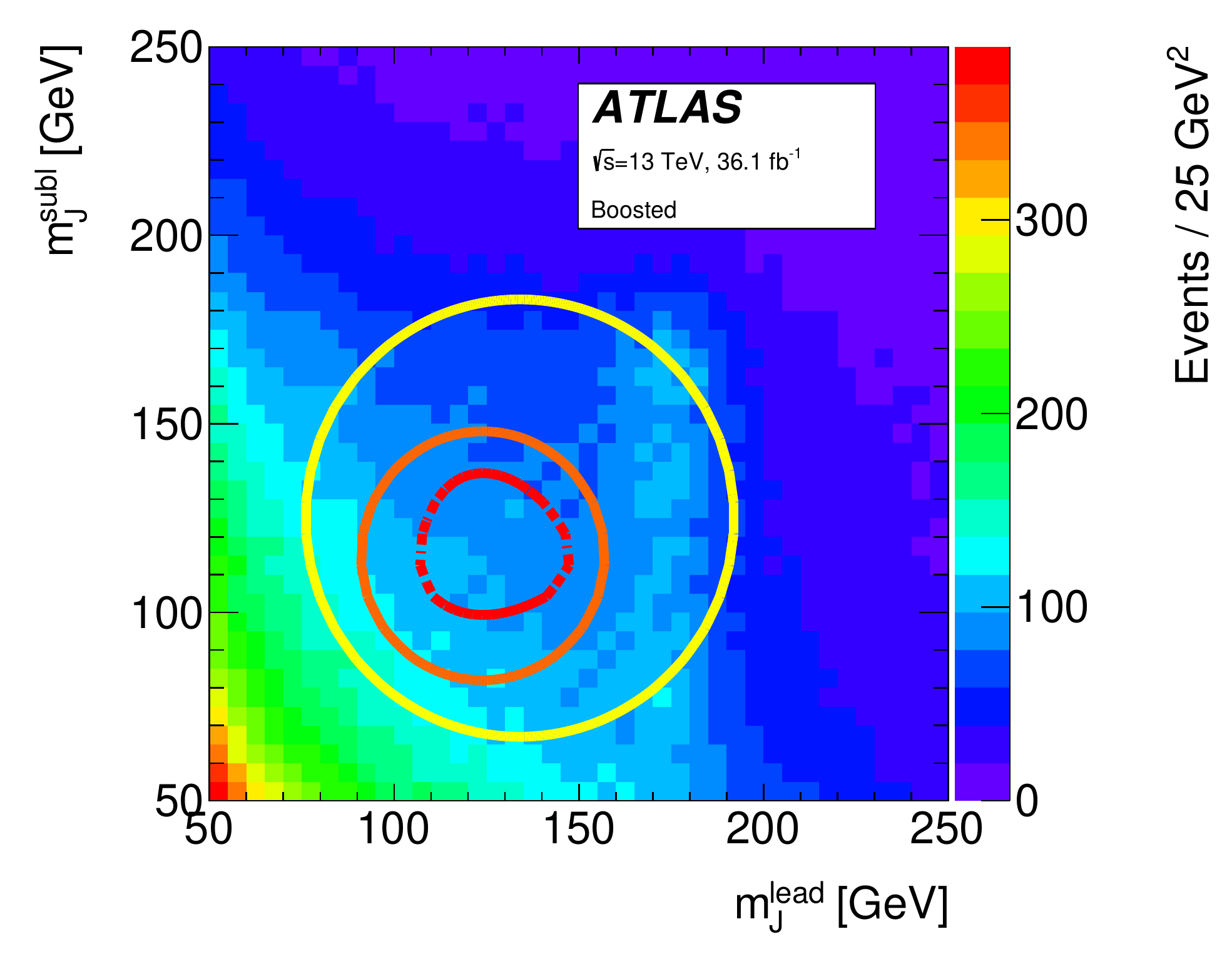}
\caption{The \mleadJ vs \msublJ distribution of the background model in the boosted two-tag sample. The signal region is the area surrounded by the inner (red) dashed contour line, centred on (\mleadJ=124~\GeV, \msublJ=115~\GeV). The control region is the area between the signal region and the intermediate (orange) contour line. The sideband region is the area between the control region and the outer (yellow) contour line.}
\label{fig:boostedRegions}
\end{center}
\end{figure}

Similarly to the resolved analysis, corrections are made for differences between the lower-tagged and $n$-tag samples by reweighting events in the lower-tagged sample. Differences between those samples are expected since requiring $b$-tags generally affects the jet kinematics. In the 1$b$-1 sample the anti-$b$-tagged large-\R jet's kinematics are reweighted to mimic the kinematics of a tagged large-\R jet (i.e. a Higgs boson candidate). Similarly, in the 2$b$-1 (2$b$-2) sample the anti-$b$-tagged large-\R jet's kinematics are reweighted to the kinematics of a Higgs boson candidate with one (two) $b$-tags. The weights are derived from data from lower-tagged samples which are inclusive in the regions, ie. events from signal, control or sideband region. Each lower-tagged sample is split into two subsamples, depending on whether the leading or sub-leading large-\R jet is $b$-tagged. The weights are obtained from spline interpolations to the ratios of the two subsamples for the three distributions that are most affected by $b$-tagging: the \pt of the leading large-$R$ jet, and the \pt of the leading track-jets associated with the leading and sub-leading large-$R$ jets. The reweighting is iterated until the weights converge to stable values.

The background yield in each of the three signal samples, $N^{n\text{-tag}}_{\mathrm{background}}$ (where $n$-tag represents two-, three- and four-tag), is evaluated using the following expression:
\begin{equation}
N^{n\text{-tag}}_{\mathrm{background}} = \muqcd^{n\text{-tag}} N^{\text{lower-tag}}_{\mathrm{multijet}} + \alphatt^{n\text{-tag}} N^{n\text{-tag}}_{t\bar{t}},
\label{eqn:boostedBackground}
\end{equation}
where $N^{\text{lower-tag}}_{\mathrm{multijet}}$ is the number of multijet events in the lower-tagged sample and $N^{n\text{-tag}}_{t\bar{t}}$ are the numbers of events predicted by the $n$-tag \ttbar\ simulation. The parameter $\muqcd^{n\text{-tag}}$ corresponds to the ratio of multijet event yields in the $n$-tag and lower-tagged samples. The parameter $\alpha_{t\bar{t}}^{n\text{-tag}}$ is a scale factor designed to correct the \ttbar\ event yield estimated from the simulation in the $n$-tag samples.

The values of $\muqcd^{n\text{-tag}}$ and $\alphatt^{n\text{-tag}}$ are extracted using Eq.~(\ref{eqn:boostedBackground}) from binned likelihood fits to the leading large-\R jet's mass distribution in the sideband region. The leading jet's mass distributions (both normalization and shape) for multijet events are obtained from the lower-tagged samples, after subtraction of the \ttbar\ contributions predicted by simulation. The fitted values of \muqcd and \alphatt are given in Table \ref{tab:boostedMuQCD}.

\begin{center}
\begin{table}[htbp!]
\caption{The fitted values of \muqcd and \alphatt for the two-tag, three-tag and four-tag samples. The statistical uncertainties are shown as well.}
\begin{center}
\begin{tabular}{l
S[table-format=1.5, table-number-alignment=left]@{\,$\pm\,$}S[table-format=1.5, table-number-alignment=left]
S[table-format=1.3, table-number-alignment=center]@{\,$\pm\,$}S[table-format=1.3, table-number-alignment=left]
} 
\toprule
Category  & \multicolumn{2}{c}{\muqcd} & \multicolumn{2}{c}{\alphatt} \\
\midrule
Two-tag 	& 0.06273 & 0.00057 & 0.986 & 0.019  \\
Three-tag & 0.1626  & 0.0043	& 0.800 & 0.073  \\
Four-tag 	& 0.0332  & 0.0043	& 0.89  & 0.60   \\
\bottomrule 
\end{tabular} 
\end{center}
\label{tab:boostedMuQCD}
\end{table}
\end{center}

The impact of limited statistical precision for $\mtwoJ > 1.2~\TeV$ in the multijet and \ttbar models is ameliorated by fitting the background distributions with the following function:
\begin{equation}
\label{eqn:dijetFn}
f\left(\mtwoJ\right) = \frac{p_1}{\left(\frac{\mtwoJ}{\sqrt{s}}\right)^2}\left(1-\frac{\mtwoJ}{\sqrt{s}}\right)^{p_2-p_3\ln{\frac{\mtwoJ}{\sqrt{s}}}},
\end{equation}
where $p_i$ are free parameters and $\sqrt{s}$ is the centre-of-mass energy. This functional form is chosen after fitting various functions to the lower-tagged data. The \mtwoJ distributions of the multijet and \ttbar backgrounds are each fitted in the range $1.2~\TeV < \mtwoJ < 3.0~\TeV$. From these fits, smooth background histograms are generated and passed to the statistical analysis. Since very few simulated \ttbar events pass the full three-tag or four-tag selections, the shape of the \ttbar distribution in the three-tag or four-tag sample is taken from the two-tag distribution. The shape differences of these templates are negligible compared to the systematic uncertainties considered and the statistical uncertainties of the available samples.

The modelling of the background yields and kinematics is validated in the control regions of the $n$-tag samples. Good agreement is observed between the yield in data and the yield of predicted backgrounds in the sideband and control regions of each of the samples, as shown in Table \ref{tab:BoostedSBCR}. Figure \ref{fig:boostedCtrl} compares the predicted background \mtwoJ\ distributions to data in the control regions of the three samples. The systematic uncertainties derived in Section \ref{sec:boostedSyst} are shown. 

\begin{table}[htbp!]
\begin{center}
\caption{The number of events in data and predicted background yields in the sideband and control regions of the two-tag, three-tag and four-tag samples for the boosted analysis. The numbers of multijet and $t\bar{t}$ background events in the sideband regions are constrained by the number of observed events, as explained in the text. The uncertainties are statistical, with fit uncertainties included for backgrounds. The anti-correlation between the multijet and \ttbar yields is accounted for in the uncertainty in the total background yield.}
\begin{center}
\begin{tabular}{l
S[table-format=5.0, table-number-alignment=right]@{\,$\pm\,$}S[table-format=3.0, table-number-alignment=left]
S[table-format=4.0, table-number-alignment=right]@{\,$\pm\,$}S[table-format=2.0, table-number-alignment=left]
S[table-format=4.0, table-number-alignment=right]@{\,$\pm\,$}S[table-format=2.0, table-number-alignment=left]
S[table-format=4.0, table-number-alignment=right]@{\,$\pm\,$}S[table-format=2.0, table-number-alignment=left]
S[table-format=3.0, table-number-alignment=right]@{\,$\pm\,$}S[table-format=2.0, table-number-alignment=left]
S[table-format=2.1, table-number-alignment=right]@{\,$\pm\,$}S[table-format=1.1, table-number-alignment=left]
} 
\toprule
  & \multicolumn{4}{c}{Two-tag} & \multicolumn{4}{c}{Three-tag} & \multicolumn{4}{c}{Four-tag} \\ 
{Source} & \multicolumn{2}{c}{Sideband} & \multicolumn{2}{c}{Control} & \multicolumn{2}{c}{Sideband} & \multicolumn{2}{c}{Control} & \multicolumn{2}{c}{Sideband} & \multicolumn{2}{c}{Control} \\ 
\midrule
Multijet & 17280 & 160  & 6848 & 67  & 3551 & 98  & 1425 & 42  & 176 & 23  & 70.4 & 8.5\\ 
\ttbar   &  7850 & 160  & 1485 & 40  & 853  & 82  & 162  & 19  & 28  & 19  & 6.4  & 4.3\\ 
Total    & 25140 & 180  & 8333 & 67  & 4404 & 77  & 1587 & 36  & 204 & 14  & 76.8 & 7.8\\ 
\midrule 
Data & \multicolumn{2}{l}{25137} & \multicolumn{2}{l}{8486} & \multicolumn{2}{l}{4403} & \multicolumn{2}{l}{1553} & \multicolumn{2}{l}{204} & \multicolumn{2}{l}{81}\\ 
\bottomrule 
\end{tabular} 
\end{center}

\label{tab:BoostedSBCR}
\end{center}
\end{table}

\begin{figure*}
\begin{center}
\subfloat[Two-tag]{\includegraphics[width=0.5\textwidth]{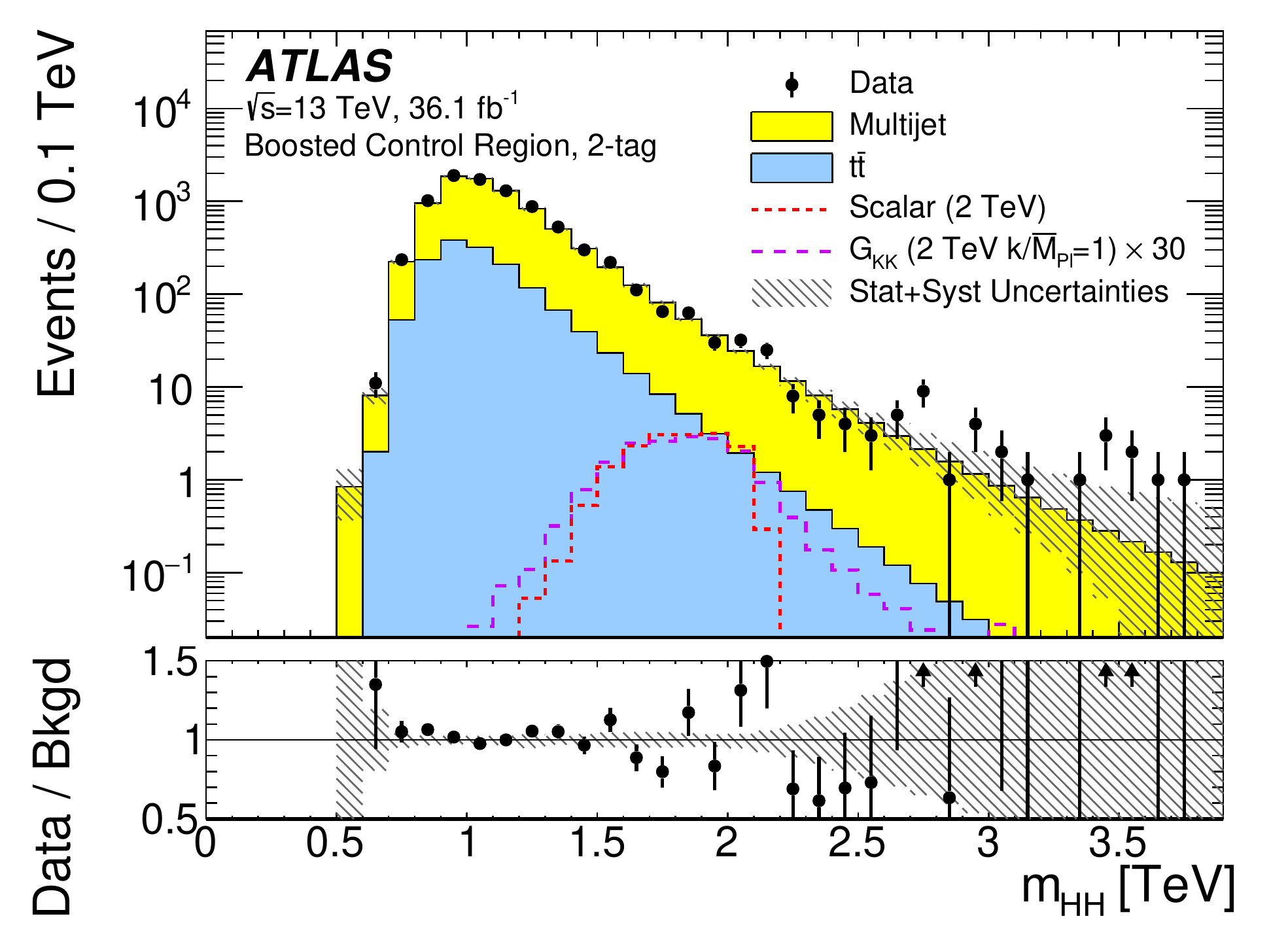}}
\subfloat[Three-tag]{\includegraphics[width=0.5\textwidth]{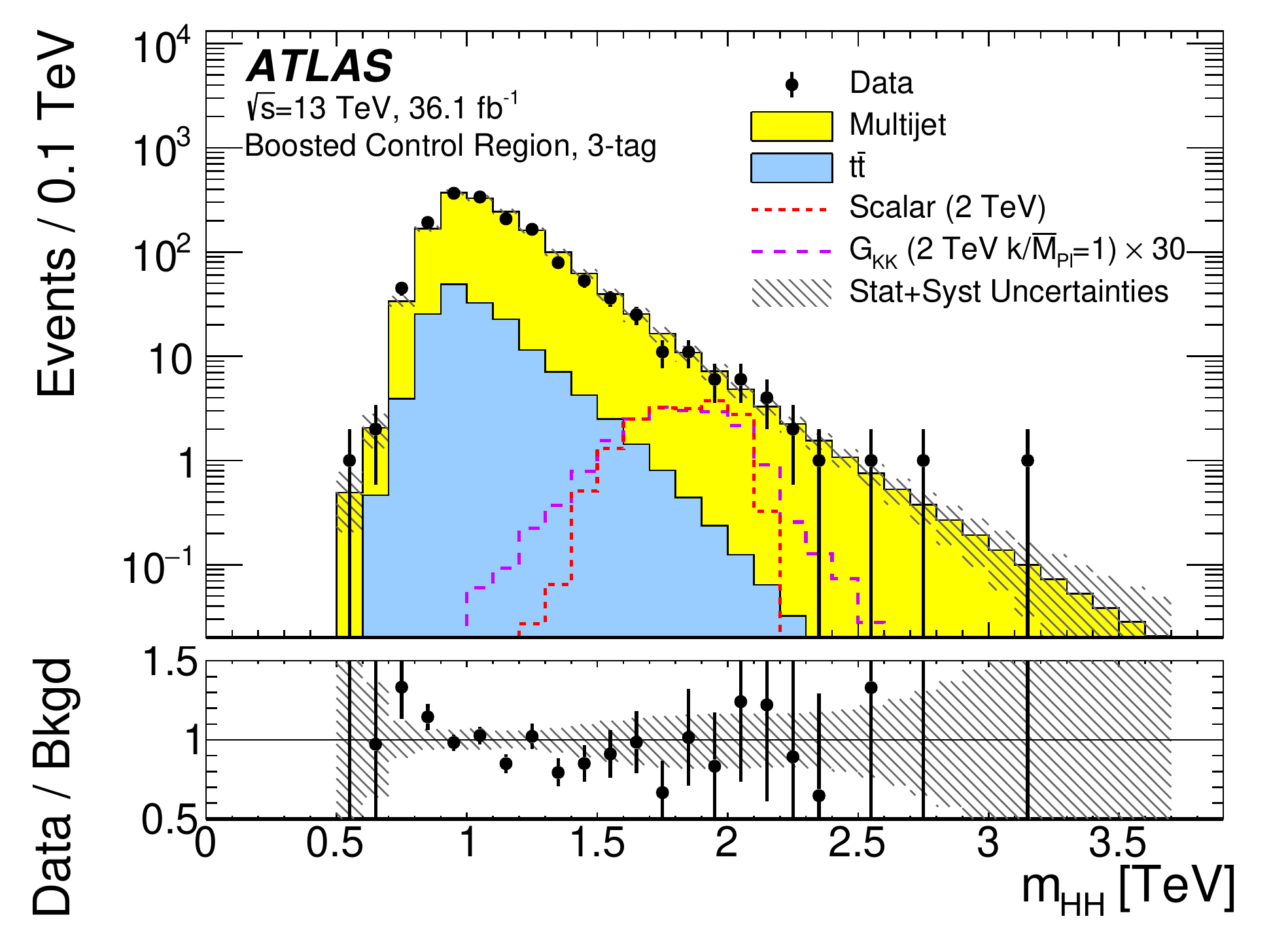}}\\
\subfloat[Four-tag]{\includegraphics[width=0.5\textwidth]{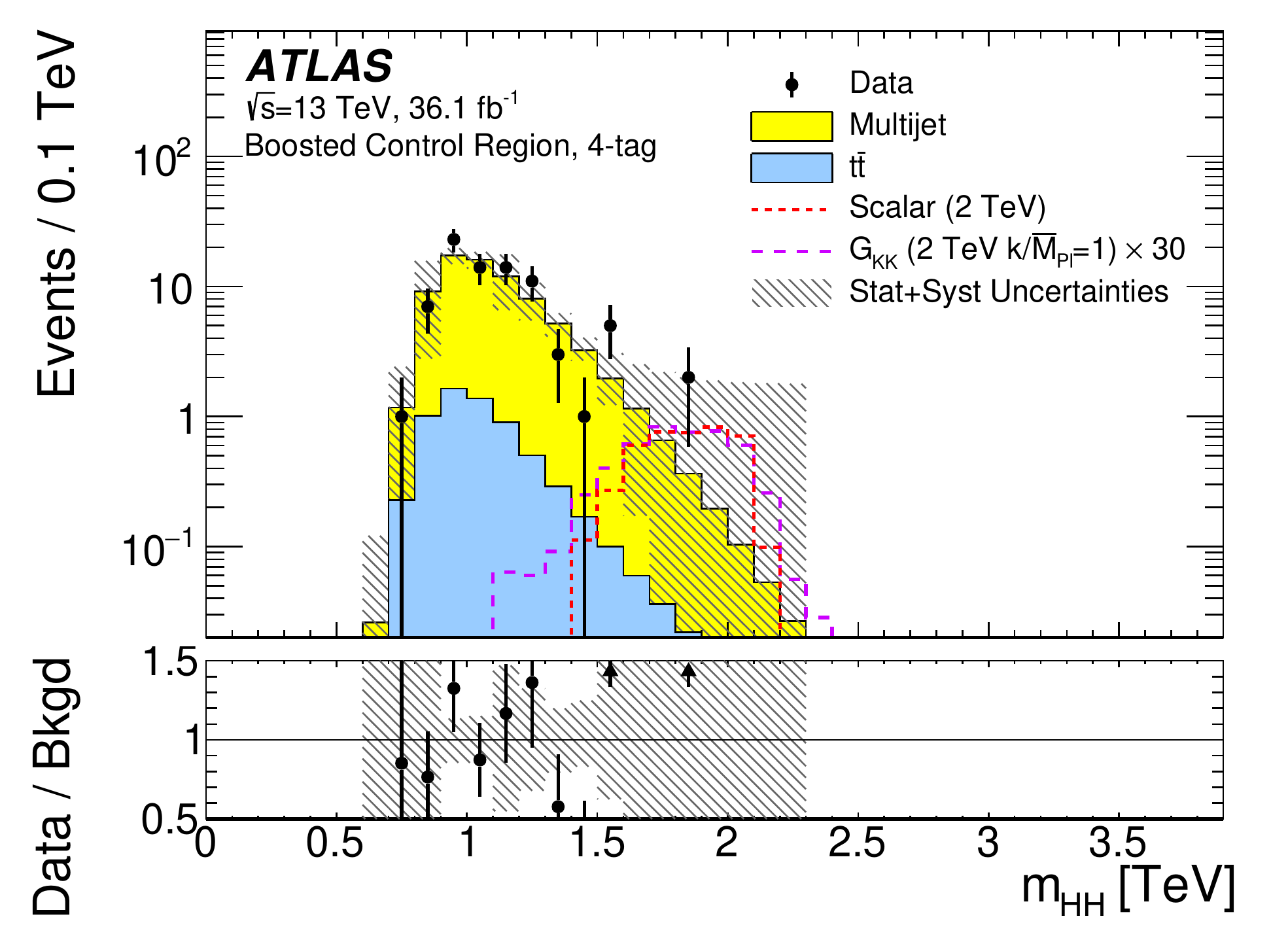}}
\caption{The \mtwoJ distributions in the control region of the boosted analysis for the data and the predicted background (top panels) for (a) the two-tag, (b) the three-tag, and (c) the four-tag samples. The data-to-background ratio (bottom panels) shows also the combination of statistical and systematic uncertainties as the grey hatched band. The expected signal for a 2~\TeV\ \Grav\ resonance with \kMPl = 1 and a scalar with the same mass is also shown. The scalar has an arbitrary cross section times branching ratio of 12~fb.}
\label{fig:boostedCtrl}
\end{center}
\end{figure*}

\subsection{Systematic uncertainties}
\label{sec:boostedSyst}

Uncertainties that are in common with those of the resolved analysis are the theoretical uncertainties in the signal acceptance and the $b$-tagging uncertainties. Systematic uncertainties that differ from those of the resolved analysis are described here.

The uncertainty in the integrated luminosity of the combined 2015 and 2016 datasets is $\pm2.1$\%~\cite{LumiCiteUP}.

The large-$R$ jet energy and mass uncertainties (i.e. jet mass scale, JMS, and jet mass resolution, JMR) are derived in situ from 13 \TeV{} $pp$ collisions, using techniques described in Ref.~\cite{JetMassAndSubstructure}. The uncertainty in the $b$-tagging efficiency for track-jets is evaluated with the same method as used for $R = 0.4$ calorimeter-based jets.

Uncertainties in the signal are treated as fully correlated across the three samples.

Detector modelling uncertainties in the \ttbar sample ($b$-tagging efficiencies; jet energy, resolution and mass) impact the result of the normalization fit. These variations of \muqcd and \alphatt  are propagated to the predictions of the multijet and \ttbar estimates in the signal regions, and they are treated as fully correlated across the three signal regions and are fully correlated with the same uncertainties in the signal.

An uncertainty in both the shape and normalization of the multijet and \ttbar backgrounds is assigned by considering the statistical uncertainties in the nominal fitted values of \alphatt and \muqcd, as given in Table~\ref{tab:boostedMuQCD}. Two orthogonal eigenvariations are calculated from the covariance matrix of the normalization fit, which are then applied to the background predictions. Correlations between \alphatt and \muqcd are fully retained this way.

An additional uncertainty in the shape of the multijet estimate, accounting for the choice of smoothing procedure, is obtained by comparing smoothed data to smoothed prediction in the control region, and assigning this difference as a shape uncertainty. The same function as defined in Eq.~(\ref{eqn:dijetFn}) is used for these smoothings, which mitigates statistical fluctuations. The uncertainty is split at $\mtwoJ = 2~\TeV$ into low- and high-\mtwoJ components.

An additional uncertainty in the normalization of the multijet background is derived from variations of the size or position of the control region or the sideband region. These variations shift the central position of the sideband and control regions by $\pm3$~\GeV\ in the directions of \mleadJ\ or \msublJ. The normalization fit to the leading large-$R$ jet's mass distribution is carried out in the varied sideband region, and the validation is done in the varied control region. For each variation, the estimated total number of background events is compared to the data, and the largest difference seen is taken as the uncertainty. The assigned uncertainties are $\pm12.2$\%, $\pm4.2$\% and $\pm2.8$\% in four-tag, three-tag and two-tag samples, respectively.

Each uncertainty in the data-driven background estimate is evaluated for each sample, and these are treated as uncorrelated across the samples in the statistical analysis.

A summary of the systematic uncertainties and their impacts on the event yields is presented in Table~\ref{tab:boostedSyst}. The impact of $b$-tagging efficiency uncertainties is smaller in the three-tag sample, since the variations applied to $b$-tagged track-jets and anti-$b$-tagged track-jets have effects that partially cancel out.

\begin{table}[htbp!]
\begin{center}
\caption{Summary of systematic uncertainties (expressed in percentage) in the total background and signal event yields in the signal region of the boosted analysis. Uncertainties are provided for each of the three samples for background, a 2~\TeV\ scalar, and a \Grav with \kMPl = 1 and $m = 2.0$~\TeV.}
\begin{center}
\resizebox{\textwidth}{!}{
\begin{tabular}{l
*9{S[table-format=2.2, table-number-alignment=center]}
} 
\toprule
  & \multicolumn{3}{c}{Two-tag} & \multicolumn{3}{c}{Three-tag} & \multicolumn{3}{c}{Four-tag} \\ 
{Source}      & {Background} & {\Grav} & {Scalar} & {Background} & {\Grav} & {Scalar} & {Background} & {\Grav} & {Scalar} \\ 
\midrule 
Luminosity    &  {-} & 2.1  & 2.1  &  {-} & 2.1   & 2.1   &  {-}  & 2.1  & 2.1 \\ 
JER           & 0.25 & 0.74 & 1    & 1.4  & 0.93  & 0.93  & 0.45  & 1.1  & 1.5 \\ 
JMR           & 0.52 & 12   & 12   & 1.4  & 12    & 13    & 7.9   & 13   & 14 \\ 
JES/JMS       & 0.43 & 1.7  & 2.1  & 2.0  & 1.9   & 2.2   & 1.3   & 3.7  & 5.7 \\ 
$b$-tagging   & 0.83 & 27   & 29   & 0.48 & 2     & 2.9   & 1.1   & 28   & 28 \\ 
Bkgd estimate & 2.8  &  {-} & {-}  & 5.8  &  {-}  & {-}   & 16    &  {-} & {-} \\ 
Statistical   & 0.6  & 1.2  & 1.3  & 1.3  & 1.0   & 1.1   & 3.1   & 1.6  & 1.9 \\ 
\midrule
Total Syst    & 3.1  & 30   & 32   & 6.6  & 13    & 14    & 18    & 31   & 32\\ 
\bottomrule
\end{tabular}
}
\end{center}
\label{tab:boostedSyst}
\end{center}
\end{table}

\subsection{Signal region event yields}
The number of events observed in the data, the predicted number of background events in the signal region, and the predicted yield for two potential signals are presented in Table~\ref{tab:boostedResults} for the two-tag, three-tag, and four-tag samples. The scalar sample shown is normalized to 12\,fb. The numbers of predicted background events and observed events are in agreement within the statistical uncertainties.

\begin{center}
\begin{table}[htbp!]
\caption{The number of predicted background events in the signal region for the boosted analysis compared to the data, for the two-tag, three-tag, and four-tag samples. The yields for a 2~\TeV~scalar and a 2~\TeV{} \Grav with \kMPl~=~1 are also shown. The scalar is normalized to a cross section times branching ratio of 12~fb. The quoted uncertainties include both the statistical and systematic uncertainties. The anti-correlation between the multijet and \ttbar yields is accounted for in the uncertainty in the total background yield.}
\begin{center}
\begin{tabular}{l
S[table-format=4.2, table-number-alignment=left]@{\,$\pm\,$}S[table-format=3.2, table-number-alignment=left]
S[table-format=3.2, table-number-alignment=center]@{\,$\pm\,$}S[table-format=2.2, table-number-alignment=left]
S[table-format=2.2, table-number-alignment=center]@{\,$\pm\,$}S[table-format=1.2, table-number-alignment=left]
} \toprule
  & \multicolumn{2}{c}{Two-tag} & \multicolumn{2}{c}{Three-tag} & \multicolumn{2}{c}{Four-tag} \\ 
\midrule
Multijet & 3390 & 150 & 702 & 63 & 32.9 & 6.9\\ 
\ttbar & 860 & 110 & 80 & 33 & 1.7 & 1.4\\ 
Total & 4250 & 130 & 782 & 51 & 34.6 & 6.1\\ 
\midrule 
\Grav (2 \TeV) & 0.97 & 0.29 & 1.23 & 0.16 & 0.40 & 0.13 \\
Scalar (2 \TeV) & 28.2 & 9.0 & 35.0 & 4.6 & 10.9 & 3.5 \\
\midrule 
Data & \multicolumn{2}{l}{4376} &  \multicolumn{2}{l}{801} &  \multicolumn{2}{l}{31}\\ 
\bottomrule 
\end{tabular}
\end{center}
\label{tab:boostedResults}
\end{table}
\end{center}

Figure~\ref{fig:boostedHHUnblinded} shows comparisons of the predicted \mtwoJ~background distributions to those observed in the data.
The predicted background and observed distributions are in agreement, with no significant excess.

\begin{figure}[!ht]
\begin{center}
\subfloat[Two-tag]{\includegraphics[width=0.5\textwidth]{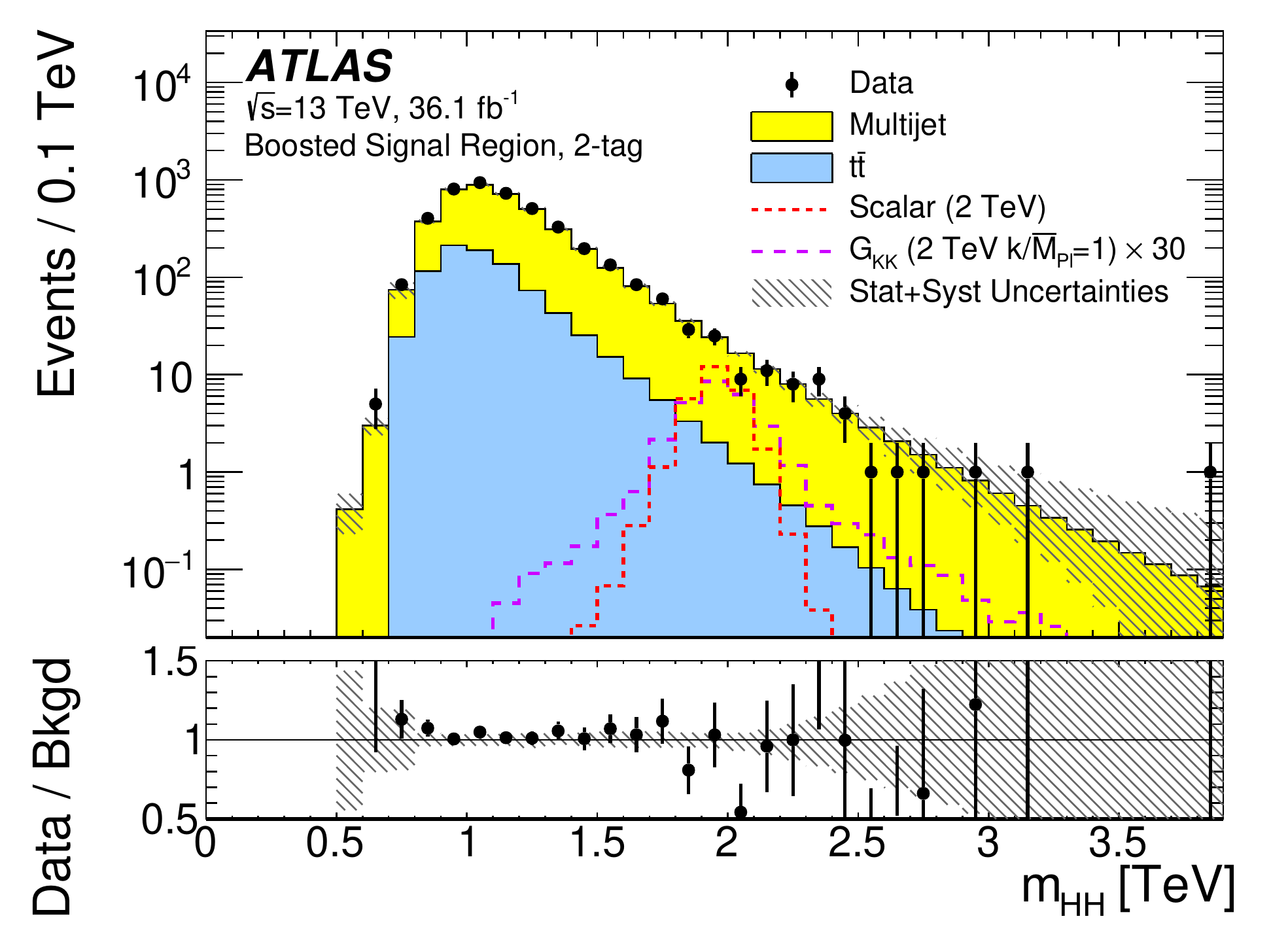}}
\subfloat[Three-tag]{\includegraphics[width=0.5\textwidth]{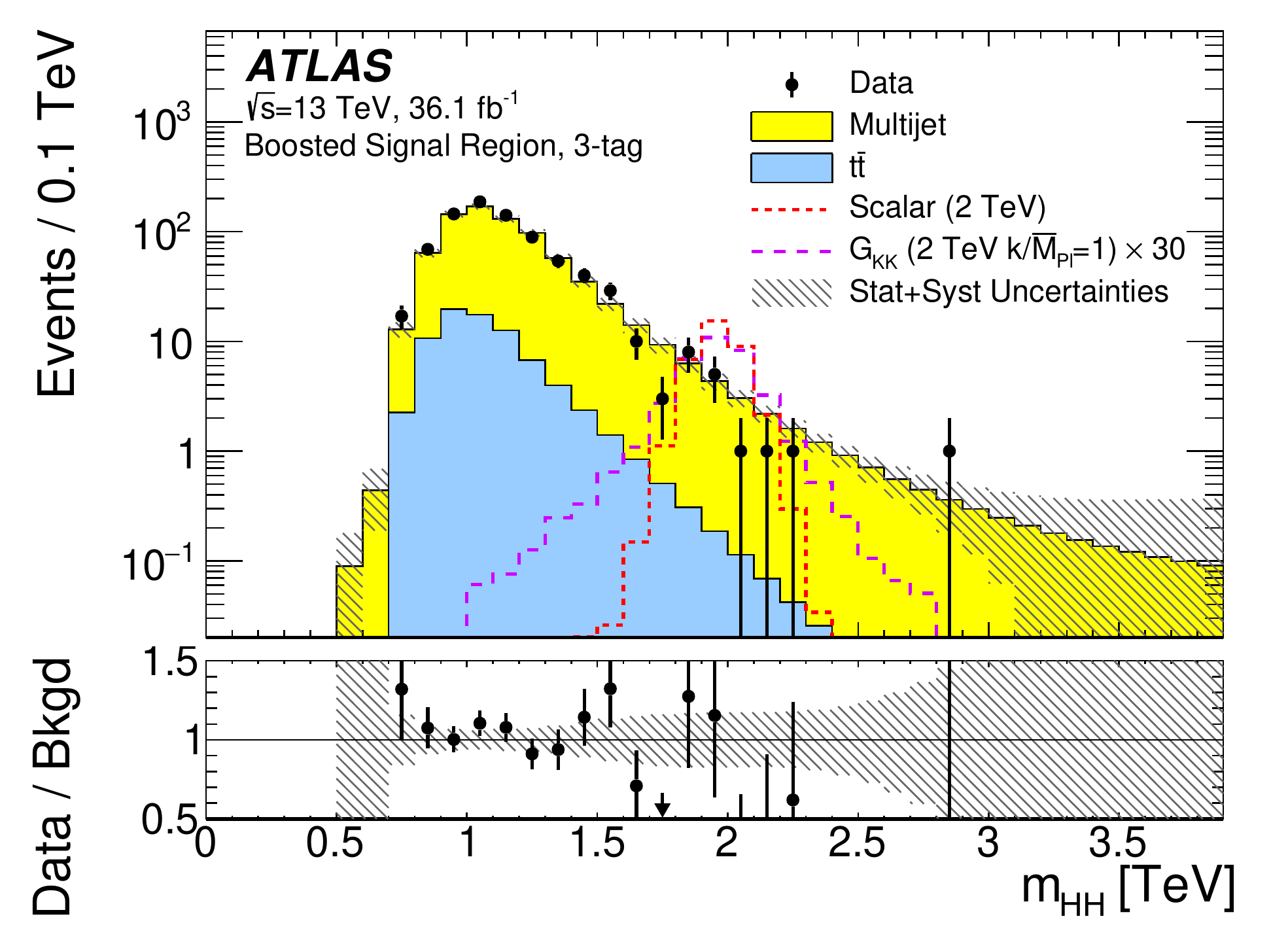}}\\
\subfloat[Four-tag]{\includegraphics[width=0.5\textwidth]{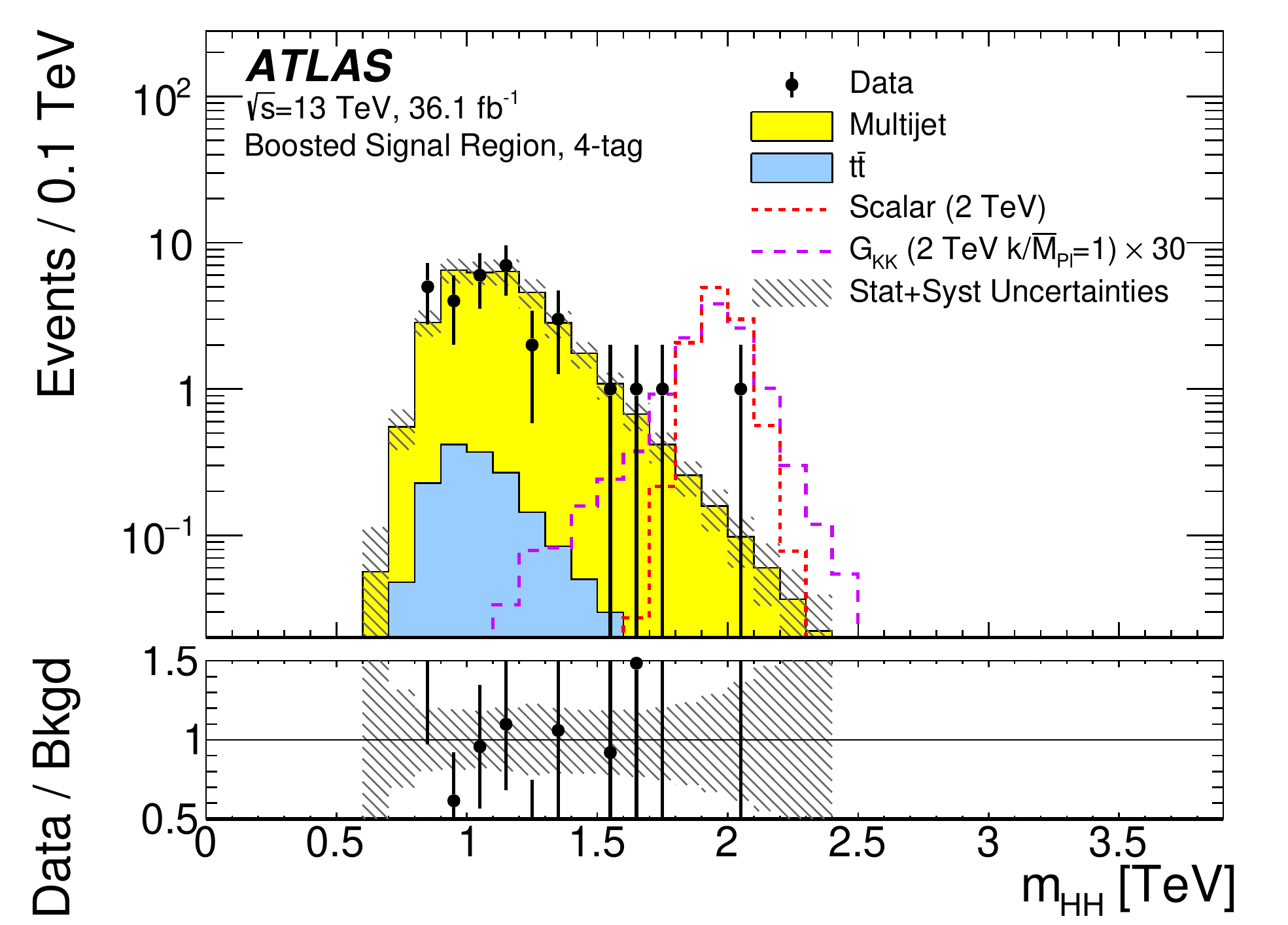}}
\caption{Distributions of \mtwoJ in the signal regions of the boosted analysis for (a) the two-tag sample, (b) the three-tag sample, and (c) the four-tag sample, compared to the predicted backgrounds. The data-to-background ratio (bottom panels) shows also the combination of statistical and systematic uncertainties as the grey hatched band. The expected signal for a 2~\TeV\ \Grav\ resonance with \kMPl = 1 and a scalar with the same mass is also shown. The scalar has an arbitrary cross section times branching ratio of 12~fb.}
\label{fig:boostedHHUnblinded}
\end{center}
\end{figure}

\FloatBarrier
\section{Statistical analysis}
\label{sec:statAna}

Following the statistical procedures outlined in Ref.~\cite{Aad:2012tfa}, a test statistic based on the profile likelihood  ratio~\cite{Cowan:2010js} is used to test hypothesized values of $\mu=\sigma/\sigma_{\mathrm{model}}$, the global signal strength factor, separately for each signal model. The exclusion limits are computed using asymptotic formulae~\cite{Cowan:2010js} and are based on the \cls method~\cite{Read:2002hq}, where a value of $\mu$ is regarded as excluded at the 95\% confidence level (CL) when \cls~is less than 5\%.

\subsection{Resonant \HH production}

The resolved analysis is performed for resonance masses in the range 260--1400~\GeV, the boosted analysis is carried out for signal masses in the range 800--3000~\GeV, and the two analyses are combined in the mass range where they overlap. The statistical analysis is performed using the data observed in the signal regions. For the resolved analysis, the \mfourj distribution is used as the final discriminant and the 2015 and 2016 datasets are fitted simultaneously. In the boosted analysis the \mtwoJ distribution is used as the final discriminant, and the data from the two-tag, three-tag and four-tag signal regions are fitted simultaneously; however, the two-tag sample is not considered for masses below 1500~\GeV\ since its contribution to the sensitivity is negligible at low mass.

Systematic uncertainties are treated within each signal region using Gaussian or log-normal constraint terms in the definition of the likelihood function. The luminosity uncertainty is treated as uncorrelated for the 2015 and 2016 datasets of the resolved analysis and the combined boosted dataset, since a subset of the 2016 dataset in the resolved analysis could not be used. All other systematic uncertainties affecting the signal are fully correlated between the resolved and boosted samples. Uncertainties in the background models are treated as uncorrelated between both analyses.

Before the fit is performed on the collision data, it is first validated on artificial datasets without statistical fluctuations. These datasets are created from the background-only model in the signal regions. The pulls, constraints, and correlations of the nuisance parameters are then checked at each mass point in the fits to the data. The impact of each uncertainty on $\mu$ is computed; the leading uncertainty at 280~\GeV\ is the background modelling uncertainty that arises from comparing the background-model derived from control region data rather than sideband data. At 2000~\GeV\ it is the background modelling uncertainty calculated by comparing smoothed data to the smoothed prediction in the control region.

Asymptotic approximations are used to calculate the local significance of a deviation from the background-only hypothesis. The largest local deviation is found at 280~\GeV, where the $p_0$ value is $1.7\cdot10^{-4}$ (3.6$\sigma$) for the narrow-width scalar, and $5.8\cdot10^{-3}$ (2.5$\sigma$) for the $\kMPl = 1$ graviton model. The $\kMPl = 2$ graviton signal is too wide to explain this deviation. The signal shape of the scalar at 280~\GeV\ has an approximately Gaussian form, resulting from the \mfourj resolution of about 9~\GeV. The graviton signals have finite widths $\Gamma$ ($\Gamma = 8~\GeV$ for $\kMPl = 1$ and $\Gamma = 33~\GeV$ for $\kMPl = 2$) and furthermore, their shapes are distorted due to their close proximity to the kinematic threshold.

The global significance is evaluated using pseudo-experiments, generated from the background-only model that has been fitted to the data. For each pseudo-experiment, the largest local excess is obtained from unconditional fits to all signal mass points and all three signal models. A distribution of those local significances is sampled, and the global $p_0$ value is computed by counting how often the largest excess in the pseudo-experiments is more significant than that observed in the data. The global significance obtained is $2.3\sigma$.

Upper limits on the cross section are set in each of the benchmark models. Figure~\ref{fig:CombinedLimits} shows the combined 95\% CL upper limits for a narrow-width scalar and a spin-2 \Grav\ in the bulk Randall--Sundrum (RS) model with $\kMPl = 1$ or $\kMPl = 2$. The predicted cross sections are shown for the graviton models, taken from Ref.~\cite{carvalho}. The bulk RS model with $\kMPl = 1$ is excluded for masses between 313 and 1362~\GeV, and the bulk RS model with $\kMPl = 2$ is excluded for masses below 1744~\GeV. The limits at low mass are stronger for the wider Graviton than the narrow Graviton model, because in the former case the signal has a larger acceptance and the distorted shape peaks at higher values of \mfourj\ above a steeply falling background.

\begin{figure}[!ht]
\begin{center}
\subfloat[Scalar]{\includegraphics[width=0.49\textwidth]{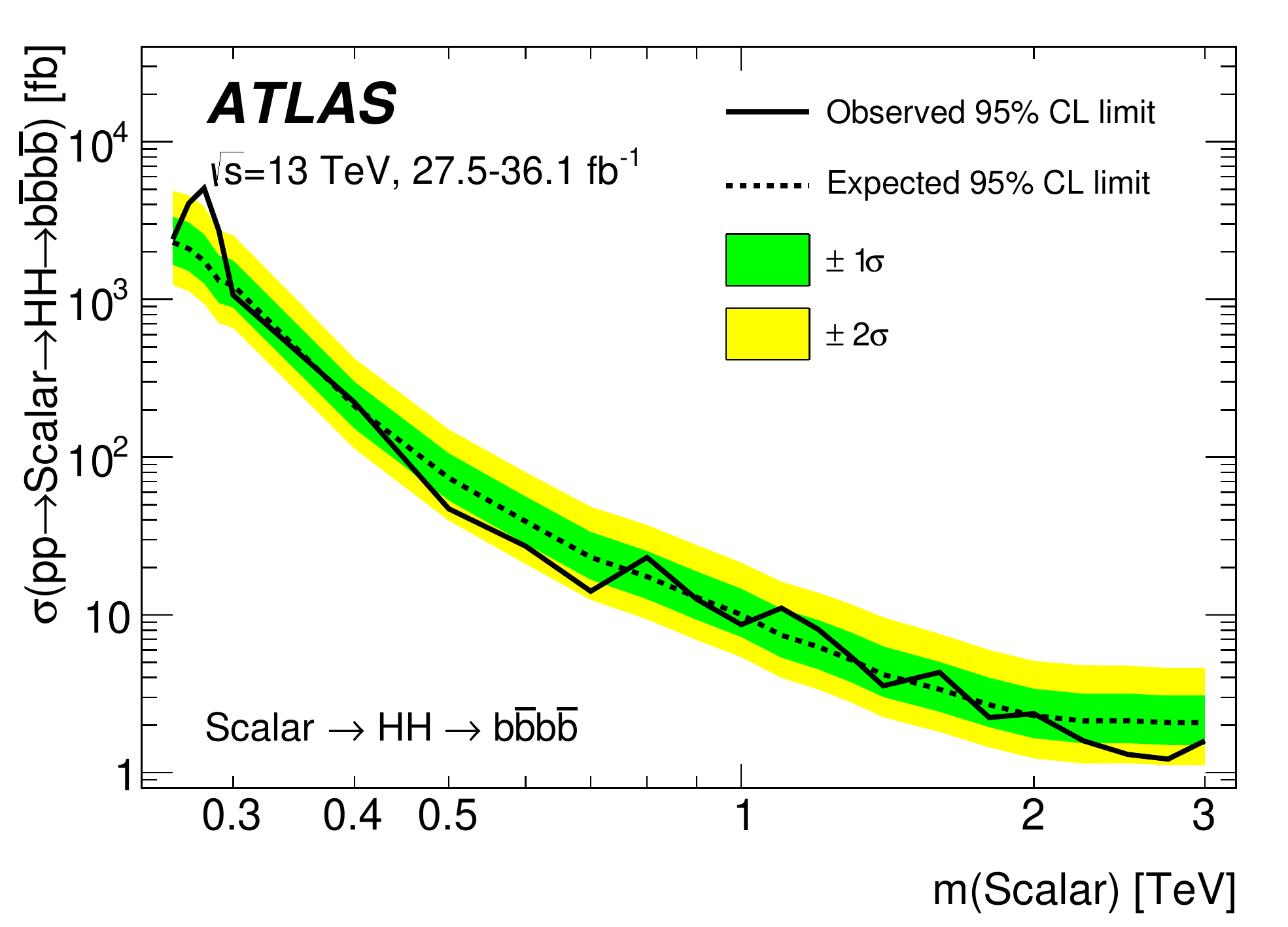}} \\
\subfloat[$\kMPl = 1$ graviton]{\includegraphics[width=0.49\textwidth]{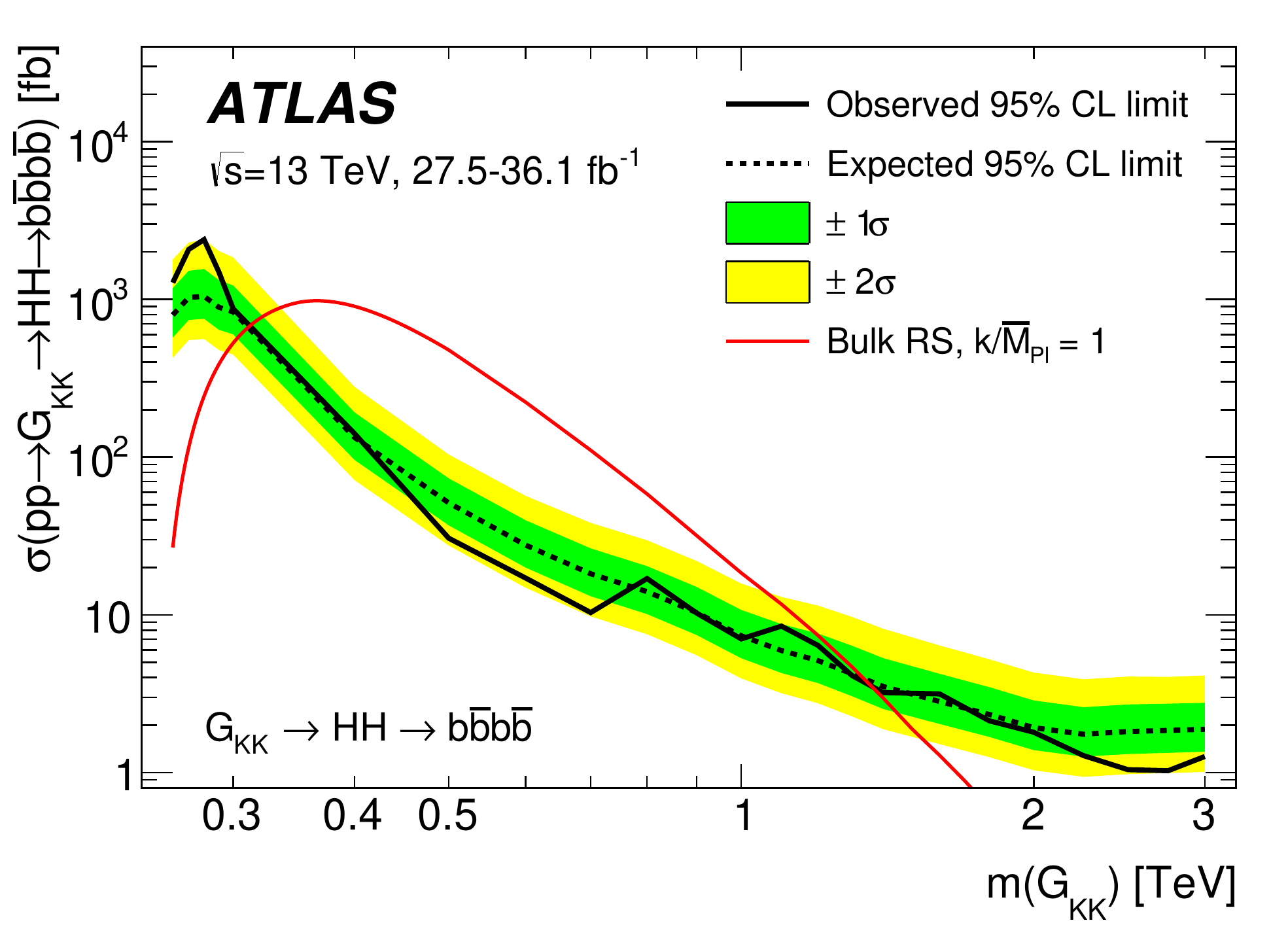}}
\subfloat[$\kMPl = 2$ graviton]{\includegraphics[width=0.49\textwidth]{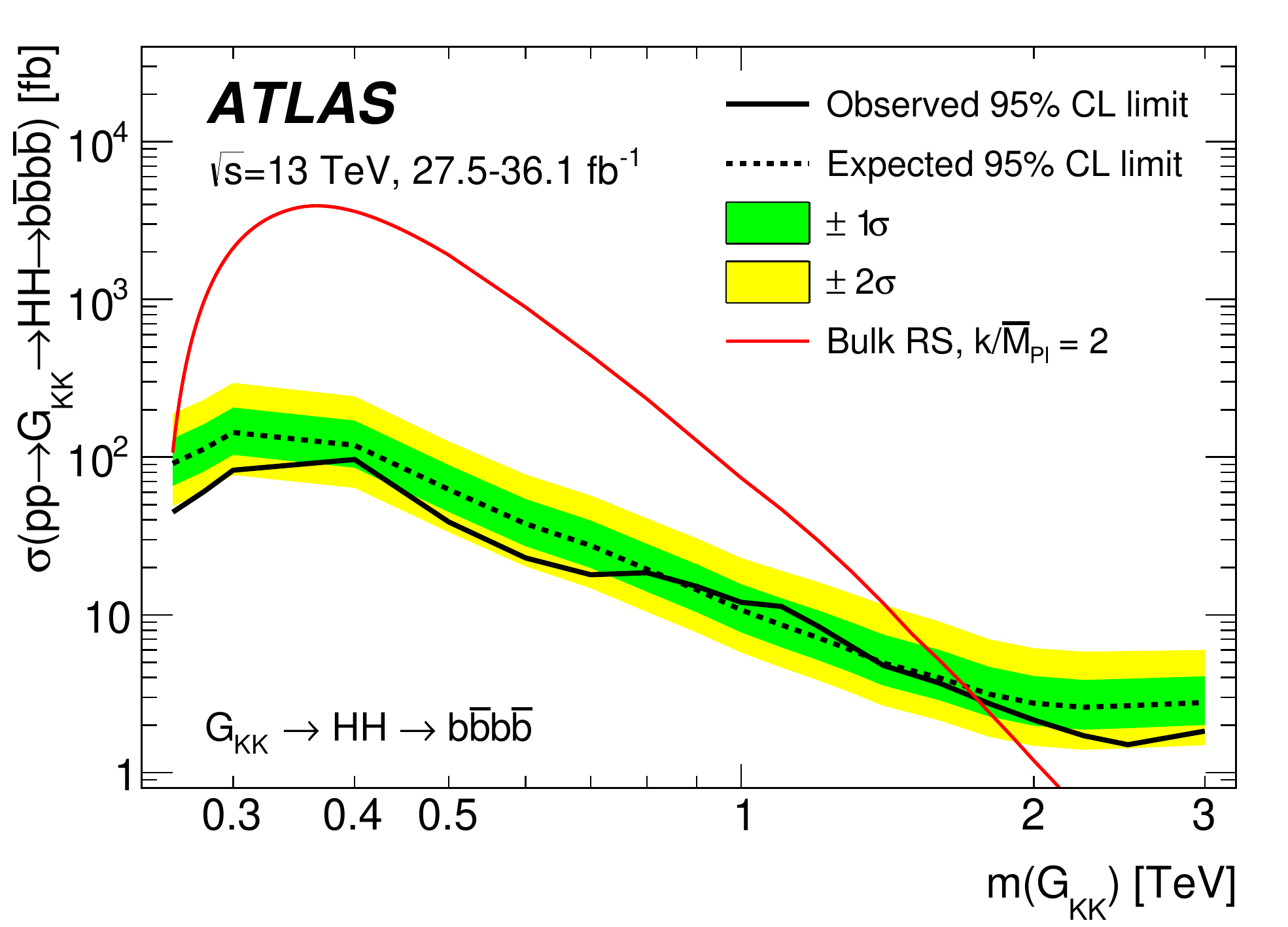}}
\caption{The observed and expected 95\% CL upper limits on the production cross section times branching ratio for the (a) narrow-width scalar, (b) bulk RS model with $\kMPl~=~1$ and (c) bulk RS model with $\kMPl~=~2$.
An additional (red) curve shows the predicted cross section as a function of resonance mass for each of the graviton models. The drop in the predicted cross sections for masses below $\mGrav = 350~\GeV$ is due to a sharp decrease in the $(\Grav\to \HH)$ branching ratio.
}
\label{fig:CombinedLimits}
\end{center}
\end{figure}

\subsection{SM non-resonant \HH production}

The non-resonant search is performed using the resolved analysis only, since it has much better sensitivity to non-resonant signals than the boosted analysis. Using the SM \HH non-resonant production via gluon--gluon fusion as the signal model, computed at NLO and fully accounting for the top-quark mass, the observed 95\% CL upper limit is $\sigma(pp\rightarrow \HH\rightarrow b\bar{b}b\bar{b}) < 147$ fb. This value is to be compared with the SM prediction for gluon--gluon fusion produced \HH of $\sigma(pp\rightarrow \HH\rightarrow b\bar{b}b\bar{b}) = 11.3^{+0.9}_{-1.0}$ fb. The expected and observed limit values and the uncertainties in the expectation are given in Table~\ref{tab:smrwMhhlims}, in units of the SM prediction. For ease of comparison with previous results, the limits are also calculated using the approximate top-quark mass correction: the observed limit is $\sigma(pp\rightarrow \HH\rightarrow b\bar{b}b\bar{b}) < 134$ fb, about 9\% smaller than that obtained with the exact top-quark mass correction.

Despite the excess at 280~\GeV, the non-resonant limits are stronger than expected. The non-resonant signal shape has its maximum around 400~\GeV, with a slowly falling spectrum towards high mass (Figure~\ref{fig:resolvedHHUnblinded}), and in several bins the number of observed events is below the prediction.

The result is limited by systematic uncertainties in the background normalization and shape. Since these are data-driven, an increase of the integrated luminosity will improve the sensitivity.

\begin{table}[htp]
\caption{95\% CL exclusion limits for SM non-resonant \HH production, in units of the SM prediction for ${\sigma(pp\rightarrow \HH\rightarrow b\bar{b}b\bar{b})}$.}
\begin{center}
\begin{tabular}{c|ccccc}
\toprule
Observed & $-2\sigma$ & $-1\sigma$ & Expected & $+1\sigma$ & $+2\sigma$ \\
\midrule
12.9 & 11.1 & 14.9 & 20.7 & 30.0 & 43.6 \\
\bottomrule
\end{tabular}
\end{center}
\label{tab:smrwMhhlims}
\end{table}

\FloatBarrier
\section{Conclusion}
\label{sec:conclusion}

A search for both resonant and non-resonant production of pairs of Standard Model Higgs bosons has been carried out in the \fourb\ channel, using 27.5--36.1~\ifb\ of LHC proton--proton collision data at $\rts = 13$~\TeV\ collected by ATLAS in 2015 and 2016.

Results are reported for the resolved analysis with each $H \to b\bar{b}$ decay reconstructed as two separate $b$-tagged small-\R jets and for the boosted analysis with each $H \to b\bar{b}$ decay reconstructed as a single large-\R jet associated with at least one $b$-tagged track-jet.

No significant data excess is observed above the estimated background consisting mainly of multijet and \ttbar\ events. The largest deviation from the background-only hypothesis is observed for narrow signal models at a mass of 280~\GeV\ in the resolved analysis, with a global significance of 2.3$\sigma$.

Upper limits on the production cross section times branching ratio to the \fourb\ final state are set for a narrow-width scalar and for spin-2 resonances with $\kMPl = 1$ or $\kMPl = 2$. The bulk RS model with $\kMPl = 1$ is excluded for masses between 313 and 1362~\GeV, and the bulk RS model with $\kMPl = 2$ is excluded for masses below 1744~\GeV. The 95\% CL upper limit on the non-resonant production is 147~fb, which corresponds to 12.9 times the SM expectation.

\clearpage

\section{Acknowledgements}

We thank CERN for the very successful operation of the LHC, as well as the
support staff from our institutions without whom ATLAS could not be
operated efficiently.

We acknowledge the support of ANPCyT, Argentina; YerPhI, Armenia; ARC, Australia; BMWFW and FWF, Austria; ANAS, Azerbaijan; SSTC, Belarus; CNPq and FAPESP, Brazil; NSERC, NRC and CFI, Canada; CERN; CONICYT, Chile; CAS, MOST and NSFC, China; COLCIENCIAS, Colombia; MSMT CR, MPO CR and VSC CR, Czech Republic; DNRF and DNSRC, Denmark; IN2P3-CNRS, CEA-DRF/IRFU, France; SRNSFG, Georgia; BMBF, HGF, and MPG, Germany; GSRT, Greece; RGC, Hong Kong SAR, China; ISF and Benoziyo Center, Israel; INFN, Italy; MEXT and JSPS, Japan; CNRST, Morocco; NWO, Netherlands; RCN, Norway; MNiSW and NCN, Poland; FCT, Portugal; MNE/IFA, Romania; MES of Russia and NRC KI, Russian Federation; JINR; MESTD, Serbia; MSSR, Slovakia; ARRS and MIZ\v{S}, Slovenia; DST/NRF, South Africa; MINECO, Spain; SRC and Wallenberg Foundation, Sweden; SERI, SNSF and Cantons of Bern and Geneva, Switzerland; MOST, Taiwan; TAEK, Turkey; STFC, United Kingdom; DOE and NSF, United States of America. In addition, individual groups and members have received support from BCKDF, CANARIE, CRC and Compute Canada, Canada; COST, ERC, ERDF, Horizon 2020, and Marie Sk{\l}odowska-Curie Actions, European Union; Investissements d' Avenir Labex and Idex, ANR, France; DFG and AvH Foundation, Germany; Herakleitos, Thales and Aristeia programmes co-financed by EU-ESF and the Greek NSRF, Greece; BSF-NSF and GIF, Israel; CERCA Programme Generalitat de Catalunya, Spain; The Royal Society and Leverhulme Trust, United Kingdom. 

The crucial computing support from all WLCG partners is acknowledged gratefully, in particular from CERN, the ATLAS Tier-1 facilities at TRIUMF (Canada), NDGF (Denmark, Norway, Sweden), CC-IN2P3 (France), KIT/GridKA (Germany), INFN-CNAF (Italy), NL-T1 (Netherlands), PIC (Spain), ASGC (Taiwan), RAL (UK) and BNL (USA), the Tier-2 facilities worldwide and large non-WLCG resource providers. Major contributors of computing resources are listed in Ref.~\cite{ATL-GEN-PUB-2016-002}.

\clearpage
\appendix

\printbibliography

\clearpage
 
\begin{flushleft}
{\Large The ATLAS Collaboration}

\bigskip

M.~Aaboud$^\textrm{\scriptsize 34d}$,    
G.~Aad$^\textrm{\scriptsize 99}$,    
B.~Abbott$^\textrm{\scriptsize 124}$,    
O.~Abdinov$^\textrm{\scriptsize 13,*}$,    
B.~Abeloos$^\textrm{\scriptsize 128}$,    
S.H.~Abidi$^\textrm{\scriptsize 165}$,    
O.S.~AbouZeid$^\textrm{\scriptsize 143}$,    
N.L.~Abraham$^\textrm{\scriptsize 153}$,    
H.~Abramowicz$^\textrm{\scriptsize 159}$,    
H.~Abreu$^\textrm{\scriptsize 158}$
% ...
\bigskip
\\

% ...
$^{13}$Institute of Physics, Azerbaijan Academy of Sciences, Baku; Azerbaijan.\\
% ...
$^{34}$$^{(a)}$Facult\'e des Sciences Ain Chock, R\'eseau Universitaire de Physique des Hautes Energies - Universit\'e Hassan II, Casablanca;$^{(b)}$Centre National de l'Energie des Sciences Techniques Nucleaires (CNESTEN), Rabat;$^{(c)}$Facult\'e des Sciences Semlalia, Universit\'e Cadi Ayyad, LPHEA-Marrakech;$^{(d)}$Facult\'e des Sciences, Universit\'e Mohamed Premier and LPTPM, Oujda;$^{(e)}$Facult\'e des sciences, Universit\'e Mohammed V, Rabat; Morocco.\\
% ...
$^{99}$CPPM, Aix-Marseille Universit\'e, CNRS/IN2P3, Marseille; France.\\
% ...
$^{124}$Homer L. Dodge Department of Physics and Astronomy, University of Oklahoma, Norman OK; United States of America.\\
% ...
$^{128}$LAL, Universit\'e Paris-Sud, CNRS/IN2P3, Universit\'e Paris-Saclay, Orsay; France.\\
% ...
$^{143}$Santa Cruz Institute for Particle Physics, University of California Santa Cruz, Santa Cruz CA; United States of America.\\
% ...
$^{153}$Department of Physics and Astronomy, University of Sussex, Brighton; United Kingdom.\\
% ...
$^{158}$Department of Physics, Technion, Israel Institute of Technology, Haifa; Israel.\\
$^{159}$Raymond and Beverly Sackler School of Physics and Astronomy, Tel Aviv University, Tel Aviv; Israel.\\
% ...
$^{165}$Department of Physics, University of Toronto, Toronto ON; Canada.\\
% ...
$^{a}$ Also at  Department of Physics, University of Malaya, Kuala Lumpur; Malaysia.\\
$^{b}$ Also at Borough of Manhattan Community College, City University of New York, NY; United States of America.\\
$^{c}$ Also at California State University, East Bay; United States of America.\\
$^{d}$ Also at Centre for High Performance Computing, CSIR Campus, Rosebank, Cape Town; South Africa.\\
$^{e}$ Also at CERN, Geneva; Switzerland.\\
% ...
$^{*}$ Deceased

\end{flushleft}


\end{document}